\documentclass[final,3p,times,twocolumn]{elsarticle}
\usepackage{graphicx}
\usepackage{color}
\usepackage[utf8]{inputenc} 
\usepackage[T1]{fontenc}    
\usepackage[colorlinks=false, hidelinks]{hyperref}
\usepackage{subcaption}
\usepackage{caption}
\usepackage{amsmath}
\usepackage{pgffor}
\usepackage{gensymb}
\usepackage{siunitx}

\usepackage{dsfont}
\usepackage{float}
\usepackage{jabbrv}
\usepackage{xcolor}
\usepackage{multirow}
\usepackage{multicol}
\usepackage{lscape}
\usepackage{bm}
\usepackage{array}
\usepackage{scalerel}
\journal{Combustion and Flame}
\usepackage{ulem}

\newcommand{\ignore}[1]{}

\begin{document}
\hypersetup{urlcolor=black,citecolor=cyan,linkcolor=black}


\begin{frontmatter}
 \title{A co-kurtosis PCA based dimensionality reduction with nonlinear reconstruction using neural networks}

\author[a]{Dibyajyoti Nayak}
 \author[a]{Anirudh Jonnalagadda}
 \author[b]{Uma Balakrishnan}
 \author[b]{Hemanth Kolla}
 \author[a]{Konduri Aditya\corref{cor1}}
 \ead{konduriadi@iisc.ac.in}
 \cortext[cor1]{Corresponding Author}
 \address[a]{Department of Computational and Data Sciences, Indian Institute of Science, Bangalore, India}
 \address[b]{Sandia National Laboratories, Livermore, California, USA}

\begin{abstract}
{
For turbulent reacting flows, identification of low-dimensional representations of the thermo-chemical state space is vitally important, primarily to significantly reduce the computational cost of device-scale simulations. Principal component analysis (PCA), and its variants, is a widely employed class of methods. Recently, an alternative technique that focuses on higher-order statistical interactions, co-kurtosis PCA (CoK-PCA), has been shown to effectively provide a low-dimensional representation by capturing the stiff chemical dynamics associated with spatiotemporally localized reaction zones. While its effectiveness has only been demonstrated based on \textit{a priori} analysis with linear reconstruction, in this work, we employ nonlinear techniques to reconstruct the full thermo-chemical state and evaluate the efficacy of CoK-PCA compared to PCA. Specifically, we combine a CoK-PCA/PCA based dimensionality reduction (encoding) with an artificial neural network (ANN) based reconstruction (decoding) and examine \textit{a priori} the reconstruction errors of the thermo-chemical state. In addition, we evaluate the errors in species production rates and heat release rates that are nonlinear functions of the reconstructed state as a measure of the overall accuracy of the dimensionality reduction technique. We employ four datasets to assess CoK-PCA/PCA coupled with ANN-based reconstruction: a homogeneous reactor for autoignition of an ethylene/air mixture that has conventional single-stage ignition kinetics, a dimethyl ether (DME)/air mixture which has two-stage ignition kinetics, a one-dimensional freely propagating premixed ethylene/air laminar flame, and a two-dimensional homogeneous charge compression ignition of ethanol. The analyses demonstrate the robustness of the CoK-PCA based low-dimensional manifold with ANN reconstruction in accurately capturing the data, specifically from the reaction zones.
}
\end{abstract}


\end{frontmatter}


\section{Introduction} \label{sec:introduction}
The multi-scale, multi-physics nature of turbulent reacting flows necessitate the use of high-fidelity simulations to accurately model chemical kinetics and turbulence-chemistry interactions. When representing chemical kinetics using first principles, e.g., direct numerical simulations with detailed kinetics, the governing system of equations has large dimensionality due to tens of chemical species participating in hundreds of chemical reactions~\cite{adityaDirectNumericalSimulation2019, savard2019, bergerDNSStudyImpact2020, Nivarti2017, desai2021direct}. As a result, the computational costs become prohibitively expensive for simulations of practical device-scale problems. Indeed, as the chemistry calculations associated with even the simplest of reaction mechanisms present themselves as the main driver of the large computational cost~\cite{uranakara2022accelerating}, reduced order modeling techniques become invaluable.

With the advent of data-driven techniques, lower-dimensional manifold (LDM) representations of the thermo-chemical subspace, identified from relevant training data, can effectively model the species dynamics of an otherwise large chemical system. Among the various available strategies to obtain these LDMs, principal component analysis (PCA) and its many flavors have been most widely employed~\cite{sutherland-parante-2009, biglari-sutherland-2012, yang-pope-chen-2013, ranade-echeckki-2019, parente-sutherland-dally-tognotti-smith-localPCA-MILD-2011, parente2009identification}. However, the principal components obtained by PCA are optimized with respect to second-order joint statistical moment, covariance, of the training data and may not be sensitive to the presence of extreme-valued samples characteristic of localized spatiotemporal events such as the formation of ignition kernels~\cite{aditya-anomaly-detection-2019-JCP}. In contrast, the statistical signature of such events is shown to be favorably captured by principal components of higher-order joint statistical moments, specifically the fourth-order co-kurtosis tensor~\cite{aditya-anomaly-detection-2019-JCP}. Building upon this observation, a dimensionality reduction procedure that constructs LDMs represented by principal components of the co-kurtosis tensor, namely the co-kurtosis PCA (CoK-PCA) method, was proposed~\cite{jonnalagadda2023co}. Additionally, analogous to PCA, a recently proposed online low-rank approximation algorithm known as dynamically bi-orthogonal decomposition (DBO), which is based on time-dependent low-dimensional subspaces, has been shown to effectively characterize strongly transient events in turbulent compressible reacting flows~\cite{chen2023direct}.

It is noteworthy that, while the CoK-PCA method was shown to represent the thermo-chemical state as well as nonlinear functions of the thermo-chemical state, such as species production rates (PRs) and heat release rates (HRRs), better than PCA in the localized spatiotemporal regions corresponding to strong chemical activity, the transformation from the principal components of the LDM to the full thermo-chemical state was performed through linear operators. However, due to the inherent nonlinear nature of the combustion phenomenon, the use of linear reconstruction has long been known not to be sufficiently accurate. Thus, the main objective of the present study is to address these concerns by studying the CoK-PCA method with nonlinear reconstruction techniques and comparing the accuracy relative to both PCA and a simple linear reconstruction.

For PCA-based LDMs, several studies have explored nonlinear reconstruction techniques such as artificial neural networks (ANNs), kernel methods, Gaussian process regression (GPR), and their hybrid approaches~\cite{mirgolbabaei2015reconstruction, echekki2015principal, malik-obando-coussement-parante-2021, 10.1007/978-3-030-80542-5_23, coussement-giquel-parante-2012}. Nonlinear reconstruction using ANN models provides flexibility to capture complex relationships, scalability for large datasets, meaningful representation learning, robustness to noise and irregularities, and the ability to generalize well to unseen data~\cite{hornik1989multilayer}. Therefore, within the confines of this paper, our primary emphasis is directed toward nonlinear reconstruction utilizing ANN. In this study, we will compare the reconstruction performance of ANNs with linear methods~\cite{jonnalagadda2023co} for thermo-chemical scalars, species production rates, and heat release rates. By contrasting the outcomes of ANN-based reconstruction with those achieved through linear techniques~\cite{jonnalagadda2023co}, we aim to evaluate the efficacy and superiority of nonlinear approaches in accurately capturing and predicting these important combustion variables. Following Jonnalagadda \textit{et al.}~\cite{jonnalagadda2023co}, the quality of the CoK-PCA-based/PCA-based encoder and ANN-based decoder models, hereafter called the CoK-PCA-ANN and PCA-ANN models, respectively, will be compared via the conventionally considered reconstruction errors of the thermo-chemical scalars as well as more sensitive PRs and HRRs for four combustion datasets namely premixed ethylene-air in a homogenous reactor, two-stage autoignition of dimethyl ether (DME)-air, a one-dimensional freely-propagating laminar flame of premixed ethylene-air, and a homogeneous charge compression ignition of ethanol-air mixture.

The remainder of this paper is organized as follows. In Sec.~\ref{sec:dimensionality-reduction}, we briefly illustrate the dimensionality reduction procedure and outline the PCA and the CoK-PCA methods to obtain low-dimensional manifolds (LDMs). Section~\ref{sec:reconstruction-methodology} describes the artificial neural network (ANN) based nonlinear reconstruction procedure to predict the thermo-chemical scalars from principal components of the LDMs. The results from \textit{a priori} analysis to evaluate the performance of the two LDMs based on ANN reconstruction are presented in Sec.~\ref{sec:results}. Finally, we summarize the paper and provide future directions in Sec.~\ref{sec:conclusions}.
\section{Dimensionality reduction}\label{sec:dimensionality-reduction}
Following convention, we arrange the scaled training data as a matrix $\mathbf{X}$~$\in$~$\mathds{R}^{(n_g\times n_v)}$ with $n_g$ observations (e.g., spatial locations, temporal checkpoints) each having $n_v$ real-valued variables or features (e.g., species concentrations, temperature). With respect to the feature space, $\mathbf{X}$ can be represented in terms of column vectors as $\mathbf{X} = \left\lbrace x_i \in \mathds{R}^{(n_g\times 1)} ~\forall ~i \in \lbrace 1, \cdots, n_v\rbrace \right\rbrace$. The purpose of dimensionality reduction, within the context of combustion, is to find a column subspace of dimension $n_q < n_v$, representing an LDM of the feature space by some measure of optimality. Note that dimensionality reduction could also denote techniques that seek an optimal row subspace, which reduces the size of $n_g$, but our interest here is strictly on reducing $n_v$. 

\subsection{Principal component analysis (PCA) based low-dimensional manifold}
For PCA, the principal vectors align in the directions of maximal variance as captured by the second order data covariance matrix, $\mathbf{C} \in \mathds{R}^{(n_v \times n_v)}$, represented using index notation as:
\begin{equation}
    (\mathbf{C})_{ij} \equiv C_{ij} = \mathds{E}(x_i x_j), \quad i,j \in \lbrace 1, \cdots, n_v\rbrace,
    \label{eq:cov_express}
\end{equation}
where $\mathds{E}$ is the expectation operator. The required principal vectors ($\mathbf{A}$) are the eigenvectors of the covariance matrix obtained through an eigenvalue decomposition, $\mathbf{C} = \mathbf{A} \mathbf{L} \mathbf{A}^T$. It should be noted that the data used in the definition of joint moments is assumed to be centered around the mean. 

\subsection{Co-kurtosis tensor based low-dimensinal manifold}
Similarly, with the higher order moment of interest, i.e., the fourth-order co-kurtosis tensor, the principal vectors represent the directions of maximal kurtosis in the data. The co-kurtosis tensor is defined as:
\begin{equation}
    T_{ijkl} = \mathds{E}(x_i x_j x_k x_l), \quad i,j,k,l \in \lbrace 1, \cdots, n_v\rbrace
    \label{eq:cok_express}
\end{equation}
By drawing an analogy to independent component analysis (ICA)~\cite{aditya-anomaly-detection-2019-JCP}, for a non-Gaussian data distribution, the fourth-order cumulant tensor, i.e., co-kurtosis $\mathbf{K}$ is computed by subtracting the excess variance given as:
\begin{equation}
        K_{ijkl} = T_{ijkl} - C_{ij} C_{kl} - C_{ik} C_{jl}- C_{il} C_{jk}
        \label{eq:fourth-order-cumulant}
\end{equation}
Again note that as the data is centered around the mean, only the second moment terms appear in the evaluation of the cumulant tensor.

The next step involves a suitable decomposition of the co-kurtosis tensor $\mathbf{K}$ to obtain the required principal components. Directly computing the higher-order joint moment tensors is expensive due to the curse of dimensionality, i.e., in our case for the co-kurtosis tensor, computational complexity would be $n_v^4$ where $n_v$ is the number of features. The symmetric nature of the co-kurtosis tensor can be leveraged to result in roughly half of $n_v^{4}$ computations. However, the existing well-defined matrix decomposition techniques cannot be directly extended to higher-order tensors. Therefore, alternate tensor decomposition methods, such as symmetric canonical polyadic (CP), higher order singular value decomposition (HOSVD), etc., should be explored to obtain the principal kurtosis vectors and values. Following~\cite{de2001independent} and~\cite{anandkumar2014}, Aditya et al.~\cite{aditya-anomaly-detection-2019-JCP} showed that the cumulant tensor $\mathbf{K}$ could be \textit{reshaped} into a $n_v \times n_v^{3}$ matrix $\mathbf{T}$ following which the principal vectors $\mathbf{U}$ are determined from the SVD of $\mathbf{T = USV}^{T}$.

After obtaining the principal components, we can reduce the dimensionality of the original data by projecting it onto a low-dimensional manifold. This is typically performed by selecting the most informative subset of principal vectors to project $\mathbf{X} \in \mathds{R}^{(n_g \times n_v)}$ onto the reduced space represented as $\mathbf{Z}_q \in \mathds{R}^{(n_g \times n_q)}$, where $n_q (<n_v)$ corresponds to the number of principal vectors retained. The conventional forward projection procedure in PCA employs a simple matrix transformation,
\begin{equation}
    \mathbf{Z}_q = \mathbf{XA}_q,
    \label{eq:forward_projection_PCA}
\end{equation}
where $\mathbf{A}_q \in \mathds{R}^{(n_v \times n_q)}$ represents the truncated subset of principal vectors (eigenvectors of the covariance matrix). For CoK-PCA, we obtain $\mathbf{A}_q$ as the $n_q$ leading left singular vectors of $\mathbf{U}$ as described above.
The contrast between PCA and CoK-PCA has been illustrated using a synthetic bivariate dataset with a few extreme-valued samples collectively representing anomalous events \cite{aditya-anomaly-detection-2019-JCP, jonnalagadda2023co}. It was observed that while the first PCA principal vector aligned in the direction of maximal variance, the first CoK-PCA principal vector aligned itself in the direction of the anomalous cluster, supporting the hypothesis that CoK-PCA is more sensitive to extreme-valued samples than PCA.

\section{Reconstruction methodology}\label{sec:reconstruction-methodology}
To assess the quality of the reduced manifold, we need to evaluate the reconstruction accuracy of the original state space from the low-dimensional subspace. Note that errors in the reconstructed variables are incurred at two stages: while projecting data into the low-dimensional space and during the reconstruction.

\subsection{Linear reconstruction}
The standard procedure of obtaining the original thermo-chemical state is a linear reconstruction through a matrix inversion, given as:
\begin{equation}
    \mathbf{X}_q = \mathbf{Z}_q\mathbf{A}_q^T,
\end{equation}
where $\mathbf{X}_q$ denotes the reconstructed data in the original feature space. Now, a comparison between $\mathbf{X}_q$ and $\mathbf{X}$ would provide a quantitative measure of the quality of the two reduced manifolds obtained by CoK-PCA and PCA, respectively. Jonnalagadda et al.~\cite{jonnalagadda2023co} analyzed the maximum and average values of the absolute reconstruction error $\left(\varepsilon = |\mathbf{X} - \mathbf{X}_q|\right)$, $\varepsilon_m = \max(\varepsilon)$ and $\varepsilon_a = \mathrm{mean}({\varepsilon})$, respectively to quantify the accuracy in each reconstructed variable. Specifically, they examined the error ratio,
\begin{equation}
r_i = \ln\left\lbrace\frac{\varepsilon_i^{\text{PCA}}}{\varepsilon_i^{\text{CoK-PCA}}}\right\rbrace,
    \label{eq:error-ratio1}
\end{equation}
to analyze the performance of CoK-PCA relative to PCA; the subscript $i$ can represent either the maximum ($r_m$) or average ($r_a$) errors.

\subsection{Nonlinear reconstruction through ANNs}
It is clear that while CoK-PCA exhibits improved accuracy in capturing stiff dynamics compared to PCA~\cite{jonnalagadda2023co}, both methods incur significant errors while employing a linear reconstruction of the original thermo-chemical state from the reduced manifold, particularly for an aggressive truncation (low $n_q$). Therefore, to fully establish the efficacy of CoK-PCA relative to PCA in capturing stiff dynamics, it is imperative to investigate its efficacy coupled with a nonlinear reconstruction approach. In this paper, we employ fully-connected deep neural networks to accomplish the required nonlinear reconstruction task. Since strong dependencies or relationships exist between different thermo-chemical scalars, it is appropriate to consider a fully-connected network where every subsequent layer is fully connected with the previous layer, ensuring the flow of information (of dependencies) across the network. In this regard, we also hypothesize that the use of a skip connection, i.e., introducing a sort of regularisation in deeper networks by skipping some of the layer outputs during backpropagation, would not be suitable. However, it should be noted that using artificial neural networks (ANNs) is an intuitive choice. Alternate nonlinear regression methods, such as Gaussian process regression (GPR), polynomial regression, least squares, etc., exist and can be incorporated in a similar manner as described in this study.

With significant advancements in deep learning in recent times, ANNs have proven their potential to model highly complex nonlinear relationships between any set of inputs and outputs. The goal of an ANN or, specifically, a deep feedforward neural network is to approximate some underlying function $f^{*}$. For example, for a classifier, $\mathbf{y} = f^{*}(\mathbf{x})$ maps an input $\mathbf{x}$ to a category $\mathbf{y}$, but more generally in case of regression problems $\mathbf{x}$ is a vector of real numbers and $\mathbf{y}$ output of a vector-valued function. A feedforward network defines a mapping $\mathbf{y} = f(\mathbf{x}; \bm{\theta})$ and learns the value of the parameters $\bm{\theta}$ that result in the best function approximation. The nonlinear reconstruction step in a dimensionality reduction algorithm can be viewed as a nonlinear mapping from the reduced manifold (or input PCs) to the original feature space (or output features). We leverage the property of ANNs being \textit{universal function approximators}~\cite{hornik1989multilayer} to achieve this task. 

Consider a reduced data representation of the original state space $\mathbf{X}$ given by the {\it score} matrix, $\mathbf{Z}_q = \mathbf{XA}_q$, where $\mathbf{A}_q \in \mathds{R}^{(n_v \times n_q)}$ comprises the chosen subset of principal vectors (kurtosis or variance). Now, the objective is to use an ANN to predict (or reconstruct) $\mathbf{X}_q$ from $\mathbf{Z}_q$ where $\mathbf{X}_q$ represents the reconstructed data in the original feature space, which is as close to $\mathbf{X}$ as possible. This is a supervised learning problem where for every $k^{th}$ feature vector from ($k^{th}$ row of) the design matrix $\mathbf{Z}_q$, $z_{k*} \in \mathds{R}^{n_q}$, the network should accurately predict the target vector ($k^{th}$ row of $\mathbf{X}$) $x_{k*} \in \mathds{R}^{n_v}$, i.e., the ANN should provide the mapping $z_{k*} \mapsto x_{k*}, ~~\forall k \in\lbrace1,2, \cdots ,n_g\rbrace$. In other words, the goal of training a neural network is to drive its prediction $\mathbf{X}_q$ to match $\mathbf{X}$. Since it is a regression problem, we evaluate the performance or accuracy of the model by using a mean squared error (MSE) loss defined as:
\begin{equation}
    \mathcal{L}_{MSE} = \dfrac{1}{m} \sum_{k=1}^{m} (\hat{x}_{k*} - x_{k*})^{2}
\end{equation}
where $\hat{x}_{k*}$, $x_{k*}$, and $m$ are the model prediction, ground truth, and the number of samples, respectively. Note that $m$ can differ from $n_g$ depending on how the entire dataset is split into training and test sets.

ANNs or feedforward networks are typically represented by composing together many different functions~\cite{goodfellow2016deep}. The model can be viewed as a directed acyclic graph describing how the functions are composed. For example, we might have three different functions $f^{(1)}, f^{(2)},$ and $f^{(3)}$ connected in a chain, to form $f(x) = f^{(3)}(f^{(2)}(f^{(1)}(x)))$. These chain-like structures form the foundation of neural networks. Each function $f^{(i)}$ corresponds to a \textit{hidden layer} of the network. The overall length of the chain gives the \textit{depth} of the network. The final layer of the network is the \textit{output layer}. Each hidden layer of the network is generally vector-valued. Every vector element can be interpreted as playing a role analogous to that of a \textit{neuron}. The dimensionality of the hidden layers (number of neurons) determines the \textit{width} of the layer. In other words, a layer can be viewed as consisting of many \textit{units (neurons)} that act in parallel, each representing a vector-to-scalar function. Each connection to a unit in a hidden layer is associated with a weight \textit{w} and a bias \textit{b}. These weights and biases parameterize the function $f^{(i)}$ for each hidden layer. The simplest feedforward neural network computes the output of a unit by a linear combination of all weights and biases associated with it. After that, a nonlinear \textit{activation} acts on this output and is responsible for inducing the required nonlinearity in the network approximation. Commonly used nonlinear activations are sigmoid, hyperbolic tangent (Tanh), rectified linear unit (ReLU), Leaky ReLU, etc. From our numerous experiments, we found that the usage of Tanh provides better stability, robustness, and smoother training of the network than ReLU, effectively handles vanishing gradients, and exhibits minimal sensitivity to different random seeds. Additionally, Tanh can map inputs to spaces with both positive and negative values, unlike ReLU and sigmoid. Thus, in this study, we employ Tanh in the hidden layers. Further, we use a custom sigmoid-based activation function at the output layer to ensure the model predictions lie within the same limits as the original state. 

Since this is a non-convex optimization problem, gradient descent-based methods are generally used to iteratively converge to the optimal solution. For our study, we use the popular Adam optimization algorithm, which is a variant of stochastic gradient descent (SGD) that realizes the benefits of two other SGD algorithms: adaptive gradient algorithm (AdaGrad) and root mean square propagation (RMSProp). Instead of using a single learning rate as in SGD, Adam computes individual adaptive learning rates for different parameters from estimates of the first and second moments of the gradients. In this case, we control the learning rate so that there is minimum oscillation when it reaches the global minimum while taking big enough steps to pass the local minimum hurdles. This method is particularly efficient for larger problem sizes involving more data or parameters. Moreover, it requires relatively lesser memory for the training procedure. The number of epochs in the training process is also selected carefully to ensure convergence.

Neural network training is inherently stochastic as it involves a random initialization of the parameters (weights and biases) at the start of the optimization. Also, the non-convexity of the loss function might result in the algorithm converging to a local minimum among multiple local minima according to a specific value of initial weights and biases. This manifests in the \textit{keras} random seed we set in our code. If the network is robust, this generally does not affect the network predictions much. Nevertheless, in this work, we employ techniques such as  stochastic weight averaging, model averaging, and ensemble averaging in the network training phase to mitigate these issues and ensure consistency in model predictions.

\subsection{Error metrics}
Once trained, the network is used to predict the thermo-chemical scalars, which include species mass fractions and temperature. The species production rates and heat release rate are also computed based on this reconstructed thermo-chemical scalars. The motivation behind calculating the species production rates and heat release rate is their nonlinear dependence on the species mass fractions and temperature, which provides a more stringent metric for assessing the reconstruction accuracy of the full thermo-chemical state and the overall dimensionality reduction strategy. Further, apart from having a tangible physical meaning, the reconstruction error associated with the heat release rate also provides an overall assessment of the quality of the reduced manifold since the heat release rate represents an aggregate effect of all the quantities of interest. A key point to note is that the network predictions correspond to a scaled version of the original state since the network is trained with scaled input feature vectors. Hence, we suitably unscale the network outputs before calculating the errors in the reconstruction of thermo-chemical scalars. Analogous to the error metrics in~\cite{jonnalagadda2023co}, we examine the following error ratios, 
\begin{equation}
r_i = \ln\left\lbrace\frac{\varepsilon_i^{\text{CoK-PCA}}}{\varepsilon_i^{\text{CoK-PCA-ANN}}}\right\rbrace,
    \label{eq:error-ratio2}
\end{equation}

\begin{equation}
r_i = \ln\left\lbrace\frac{\varepsilon_i^{\text{PCA}}}{\varepsilon_i^{\text{PCA-ANN}}}\right\rbrace,
    \label{eq:error-ratio3}
\end{equation}

\begin{equation}
r_i = \ln\left\lbrace\frac{\varepsilon_i^{\text{PCA-ANN}}}{\varepsilon_i^{\text{CoK-PCA-ANN}}}\right\rbrace,
    \label{eq:error-ratio4}
\end{equation}
to compare the relative performance of different methods such as CoK-PCA, PCA, CoK-PCA-ANN, and PCA-ANN considered in our study. Again, the subscript $i$ can represent either the maximum ($m$) or average ($a$) errors. The value of $r_i$ will be positive if the ratio inside the logarithm is greater than unity (the error in the denominator is lower), indicating that the technique represented by the denominator is more accurate than that represented by the numerator. In the results to be shown, following \cite{jonnalagadda2023co}, we will denote positive $r_i$ by blue and negative by brown colored bars.
\section{Results}\label{sec:results} 
To investigate the accuracy of the proposed reconstruction methodology for combustion datasets, we consider four test cases representative of various physical and chemical phenomena (e.g., autoignition, flame propagation) ubiquitous in such scenarios:
\begin{itemize}
\item autoignition of a premixed ethylene/air mixture in a homogeneous reactor,
\item autoignition, with two-stage ignition kinetics, of a dimethyl ether (DME)/air mixture in a homogeneous reactor,
\item one-dimensional freely propagating planar laminar premixed flame of ethylene/air mixture,
\item two-dimensional turbulent autoignition of ethanol/air at homogeneous charge compression ignition (HCCI) conditions.
\end{itemize}
The datasets represent an increasing order of complexity of chemical kinetics and flow-chemistry interactions. The first two cases represent homogeneous (spatially zero-dimensional) autoignition, albeit ethylene/air with conventional ignition kinetics, while DME/air has more complex low and high temperature ignition kinetics. The third case incorporates spatial variation, including convection and diffusion effects in the canonical planar laminar premixed flame configuration. The fourth case represents complex turbulence-chemistry interactions in a spatially 2-D configuration under conditions relevant to practical devices. 
\subsection{Premixed ethylene-air in a homogeneous reactor}\label{sec:0D_ethylene_air}
\begin{figure}[h!]
    \centering
    \includegraphics[width = 4.5cm]{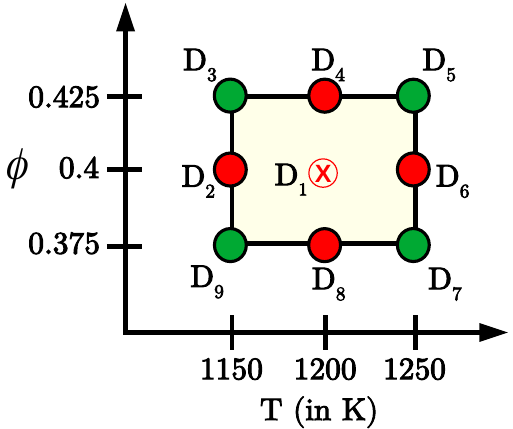}
    \caption{Illustration of train-test split in ensemble training. Training states: $D_1, D_2, D_4, D_6, D_8$ and testing states: $D_3, D_5, D_7, D_9$. To generate the LDMs, PCs are computed based on the reference state, $D_1$. 
    }
    \label{fig:0D_9f_ensemble_training}
\end{figure}
In this section, we consider the dataset that characterizes spontaneous ignition in a simple homogeneous (zero-dimensional) reactor. For dataset generation, we simulate a constant pressure reactor with a premixed ethylene-air mixture at a pressure P = \SI{1.72}{atm} for a suite of nine flamelets, i.e., $D_i~ \forall i \in \lbrace 1, 2, \cdots, 9 \rbrace$, each with a different initial temperature (T) and equivalence ratio ($\phi$) as illustrated in Fig.~\ref{fig:0D_9f_ensemble_training}. Specifically, we perturb the initial conditions ($\mathrm{T}, \phi$) from a reference state of $D_1 \equiv$ ($\mathrm{T} = \SI{1200}{K}$, $\phi = 0.4$) by $\Delta T = \pm \SI{50}{K}$  and $\Delta \phi = \pm 0.25$. Thus, each flamelet is parameterized by a combination of initial (T, $\phi$) where T~$\in \lbrace \SI{1150}{K}, \SI{1200}{K}, \SI{1250}{K} \rbrace$ and $\phi~\in \lbrace 0.375, 0.4, 0.425 \rbrace$. The chemistry is represented by a 32-species, 206-reactions mechanism~\cite{luo2012chemical}. The homogeneous reactor simulations are performed with Cantera~\cite{Cantera}, and each flamelet is computed for different durations to ensure that the profiles remain nearly similar. For the reference state, the reactor is evolved for \SI{2.5}{\milli\second} with a time step of \SI{1}{\micro \second} to yield 2501 data samples. Hence, in this case, the original design matrix \textbf{D} consists of $n_g$ = 2501 points and $n_v$ = 33 variables, comprising 32 species and temperature. The next step involves a data preprocessing stage where the design matrix for each state is zero-centered by subtracting with the mean feature vector and normalized with the absolute maximum feature vector to obtain the scaled data matrix, $\mathbf{X}$. This ensures an unbiased data representation with equal weightage given to all the features. To generate the low-dimensional manifolds, i.e., using PCA and CoK-PCA, we compute the principal vectors and values based on the scaled reference state ($X_1$), which eventually forms the basis for constructing the training/validation data. Next, we perform an aggressive truncation of the reduced manifolds by retaining $n_q = 5$ dominant principal vectors out of the $n_v = 33$ vectors that capture approximately 99\% of the variance and 98\% of the kurtosis in the dataset, respectively. Using the principal vectors computed on the scaled reference state ($X_1$), we obtain the LDM representation (score matrices) $\mathbf{Z}_{q}^{4}$ and $\mathbf{Z}_{q}^{2}$ through the dimensionality reduction procedure discussed in Sec.~\ref{sec:dimensionality-reduction} for the CoK-PCA and PCA reduced manifolds, respectively. It should be noted that this projection is a linear operation.
\begin{figure}[h!]
    \centering
    {
    \includegraphics[width = 6cm]{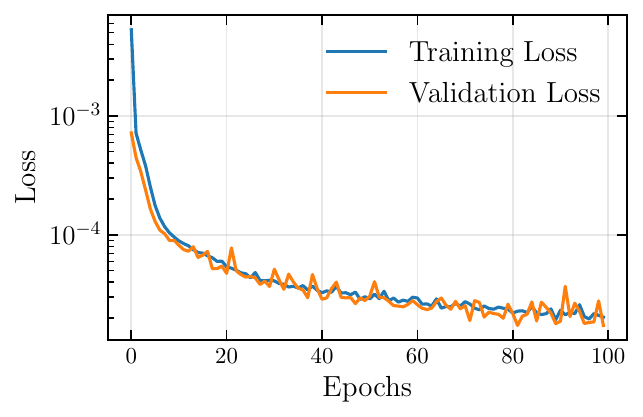}
    \begin{picture}(0,0)
        \put(-130,90){\scriptsize (a)}
    \end{picture}
    }
    {
    \includegraphics[width = 6cm]{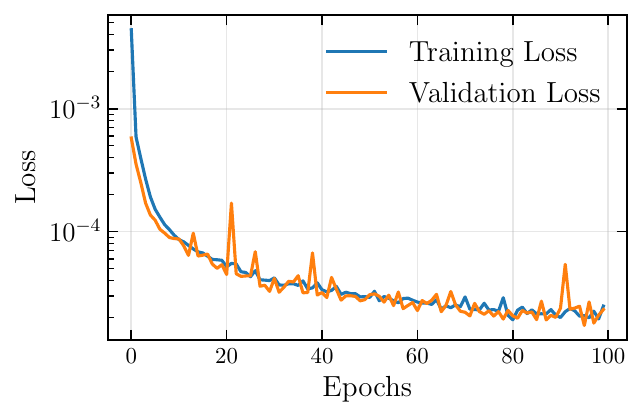}
    \begin{picture}(0,0)
        \put(-130,90){\scriptsize (b)}
    \end{picture}
    }
    \caption{Training and validation loss curves for (a) CoK-PCA-ANN and (b) PCA-ANN, respectively, for the premixed ethylene-air homogeneous reactor dataset.}
    \label{fig:0D_9f_loss_curves}
\end{figure}
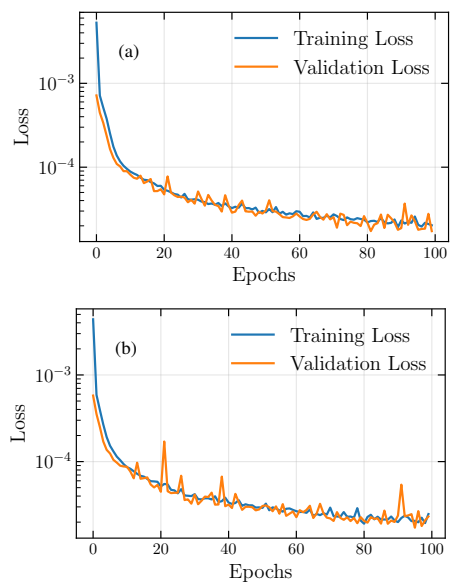

After obtaining the LDMs with PCA and CoK-PCA, the next step in the \textit{a priori} analysis is to evaluate the reduced manifolds in conjunction with the nonlinear reconstruction of the original thermo-chemical state through ANNs. For the ANN training phase, the input feature vectors are the rows of the score matrices ($\mathbf{Z}^{4}_{q}, \mathbf{Z}^{2}_{q}$) and output vectors are the corresponding rows of the scaled original thermo-chemical state matrix $\mathbf{X}$; these matrices are arranged based on the different flamelets ($D_j$s) using train-test split shown in Fig.~\ref{fig:0D_9f_ensemble_training}, i.e., flamelets $D_1$, $D_2$, $D_4$, $D_6$, and $D_8$ are used for ANN training only. Through hyperparameter tuning, the best network architecture is ascertained with four hidden layers of widths of 40, 64, 40, and 32 neurons, respectively. In addition, the widths of input and output layers correspond to $n_q = 5$ and $n_v = 33$ neurons, respectively. Further, we use a hyperbolic tangent activation in the hidden layers and a custom sigmoid-based activation at the output layer, which ensures the network predictions are bounded in the same limits as the scaled inputs. Finally, we employ the widely used Adam optimizer (learning rate = \num{1e-3}) to facilitate robust, stable, fast network learning. Figure~\ref{fig:0D_9f_loss_curves} shows the loss curves obtained for CoK-PCA-ANN and PCA-ANN, where convergence is achieved at around 100 epochs with a validation loss of about \num{2e-5}.

Having trained on a subset of the flamelets, we use the neural network to predict (or reconstruct) the scaled species mass fractions and temperature for the test states, i.e., $D_j~\forall j \in \lbrace 3,5,7,9 \rbrace$. To ensure that the reconstructed thermo-chemical state results in a unit sum of species mass fractions, as is the standard practice, all reconstructed species mass fractions which yield negative values (that are slightly smaller than zero) are taken to be zero, after which any deviation from the sum equalling unity is adjusted for in the non-participating or bath species. Using the reconstructed thermo-chemical scalars, $\mathbf{D_q}$, we proceed to compute the species production rates and heat release rates. The reconstructed quantities are compared against the original thermo-chemical state, $\mathbf{D}$, and their derived quantities (species production rates, heat release rates) using the error metrics, $r_m$ and $r_a$.
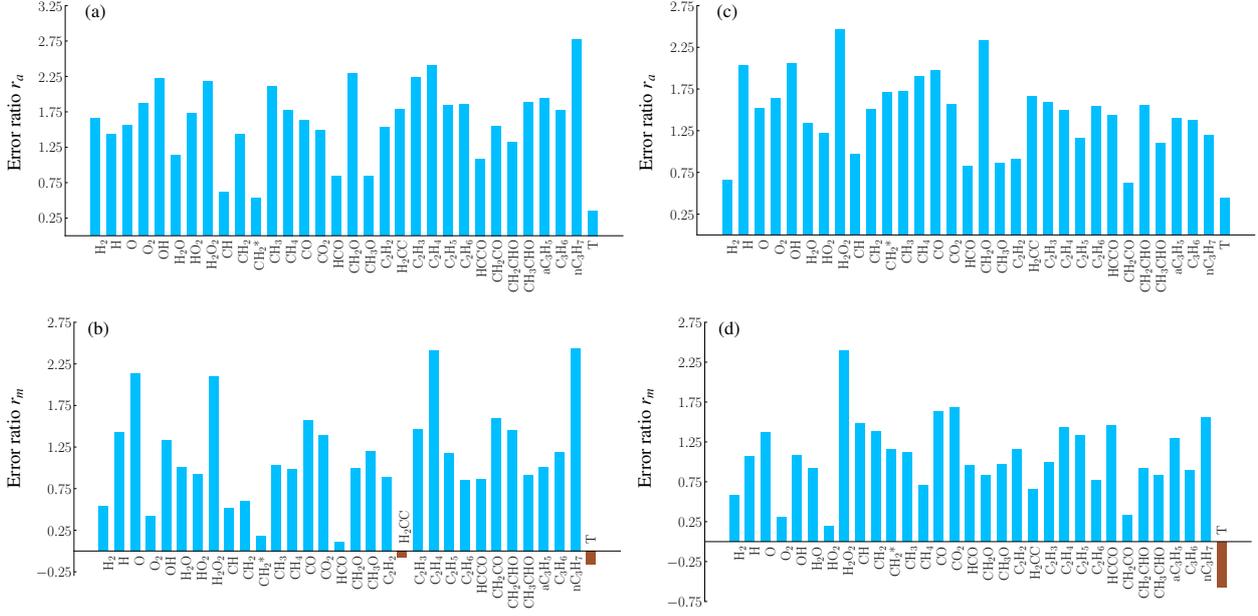
\begin{figure*}[h!]
    \centering
    {
    \includegraphics[width = 7.98cm]{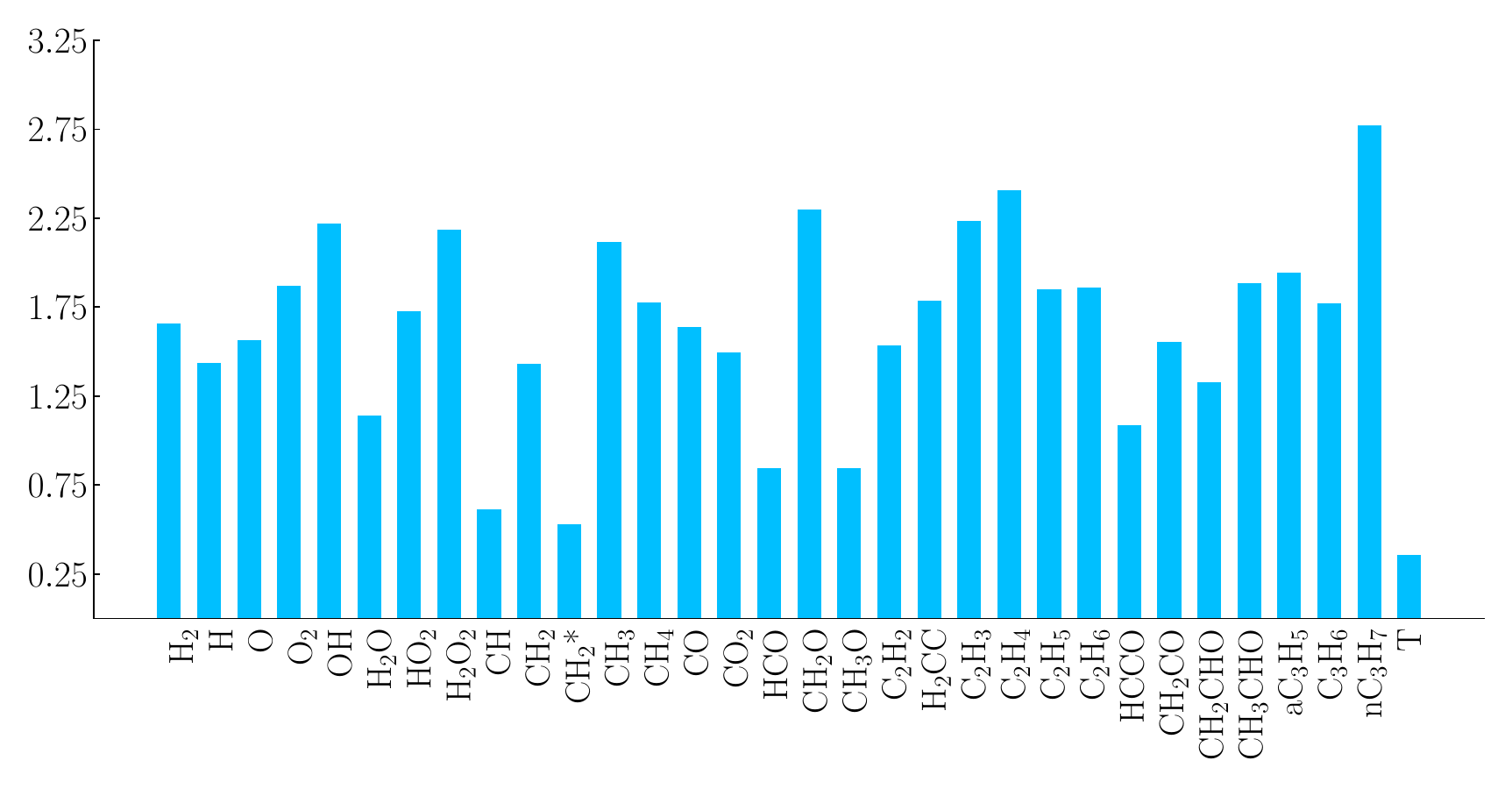}
    \begin{picture}(0,0)
        \put(-208,108){\scriptsize(a)}
        \put(-237,50){\scriptsize {\rotatebox{90}{Error ratio $r_a$}}}
    \end{picture}
    }\hspace{-0.018cm}
    {
    \includegraphics[width=7.98cm]{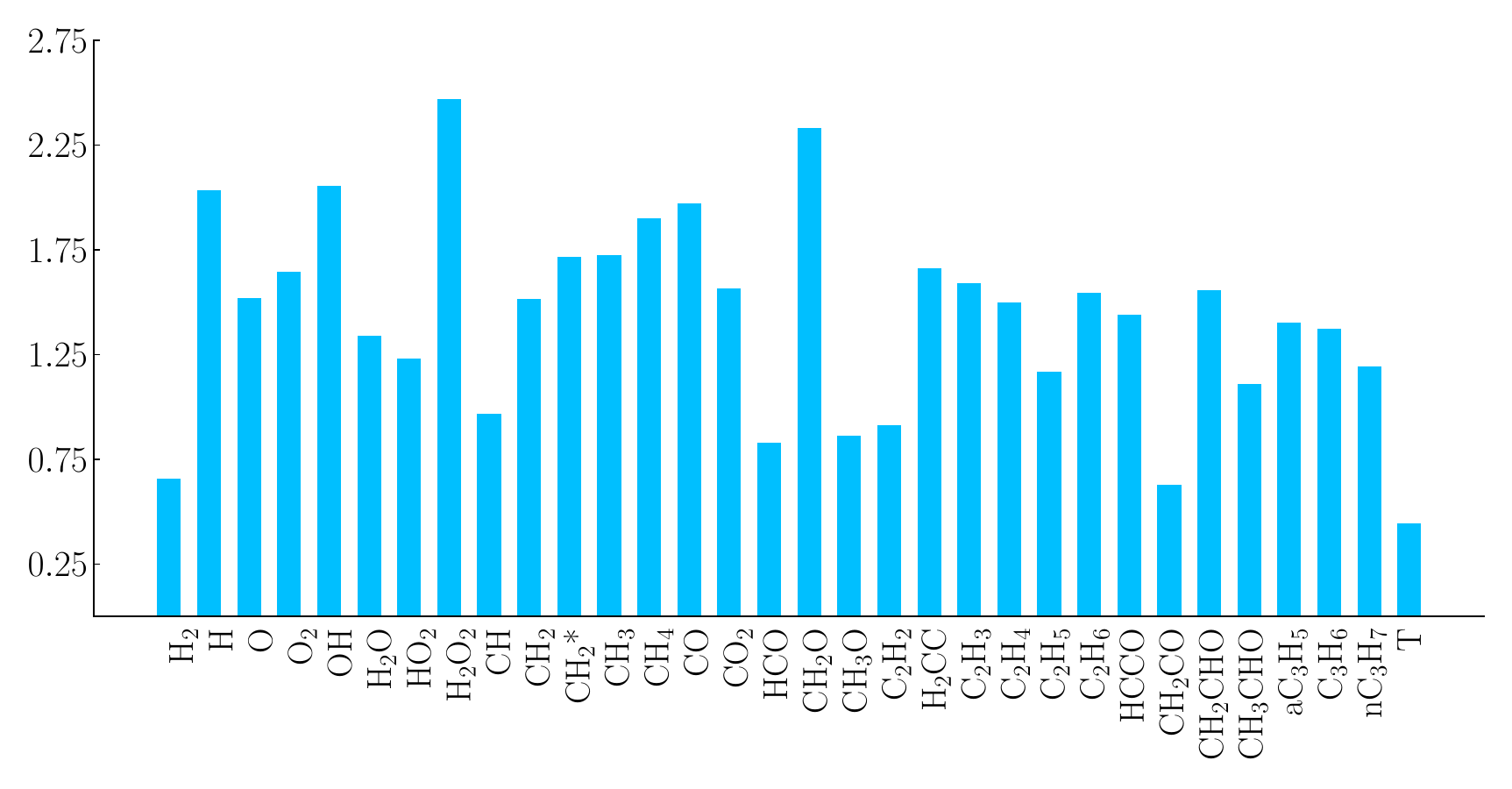}
    \begin{picture}(0,0)
        \put(-208,108){\scriptsize(c)}
        \put(-237,50){\scriptsize {\rotatebox{90}{Error ratio $r_a$}}}
    \end{picture}
    }
    {
    \includegraphics[width=7.98cm]{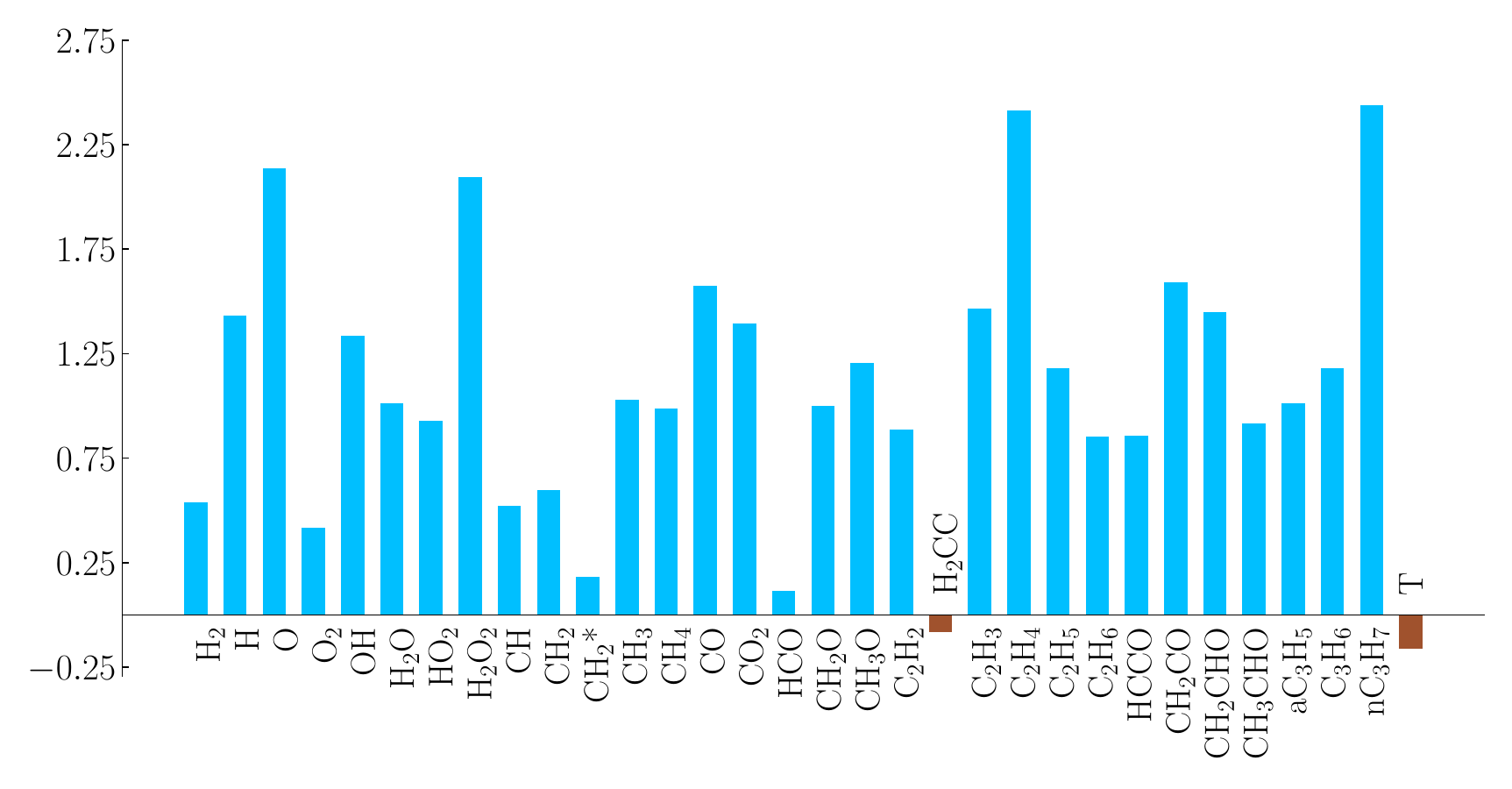}
    \begin{picture}(0,0)
        \put(-206,108){\scriptsize(b)}
        \put(-236,50){\scriptsize {\rotatebox{90}{Error ratio $r_m$}}}
    \end{picture}
    }\hspace{-0.018cm}
    {
    \includegraphics[width=7.98cm]{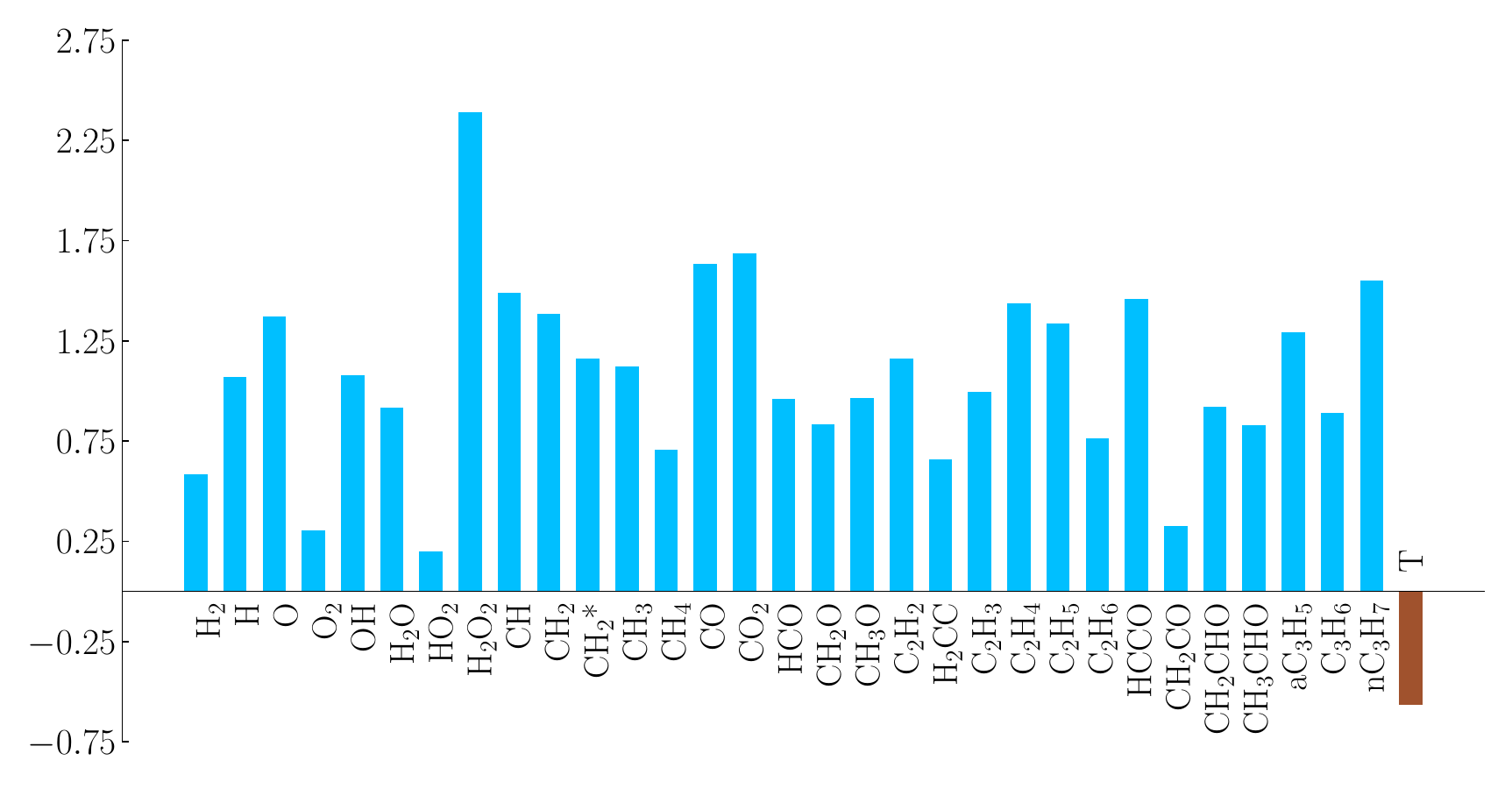}
    \begin{picture}(0,0)
        \put(-206,108){\scriptsize(d)}
        \put(-236,50){\scriptsize {\rotatebox{90}{Error ratio $r_m$}}}
    \end{picture}
    }
    \caption{Comparison of errors in the reconstruction of thermo-chemical scalars for (a), (b) CoK-PCA vs. CoK-PCA-ANN and (c), (d) PCA vs. PCA-ANN for the premixed ethylene-air homogeneous reactor dataset. Top and bottom plots in each column represent $r_a$ and $r_m$ respectively.}
    \label{fig:0D_9f_tsc_linear_nonlinear}
\end{figure*}

\begin{figure*}[h!]
    \centering
    {
    \includegraphics[width = 7.98cm]{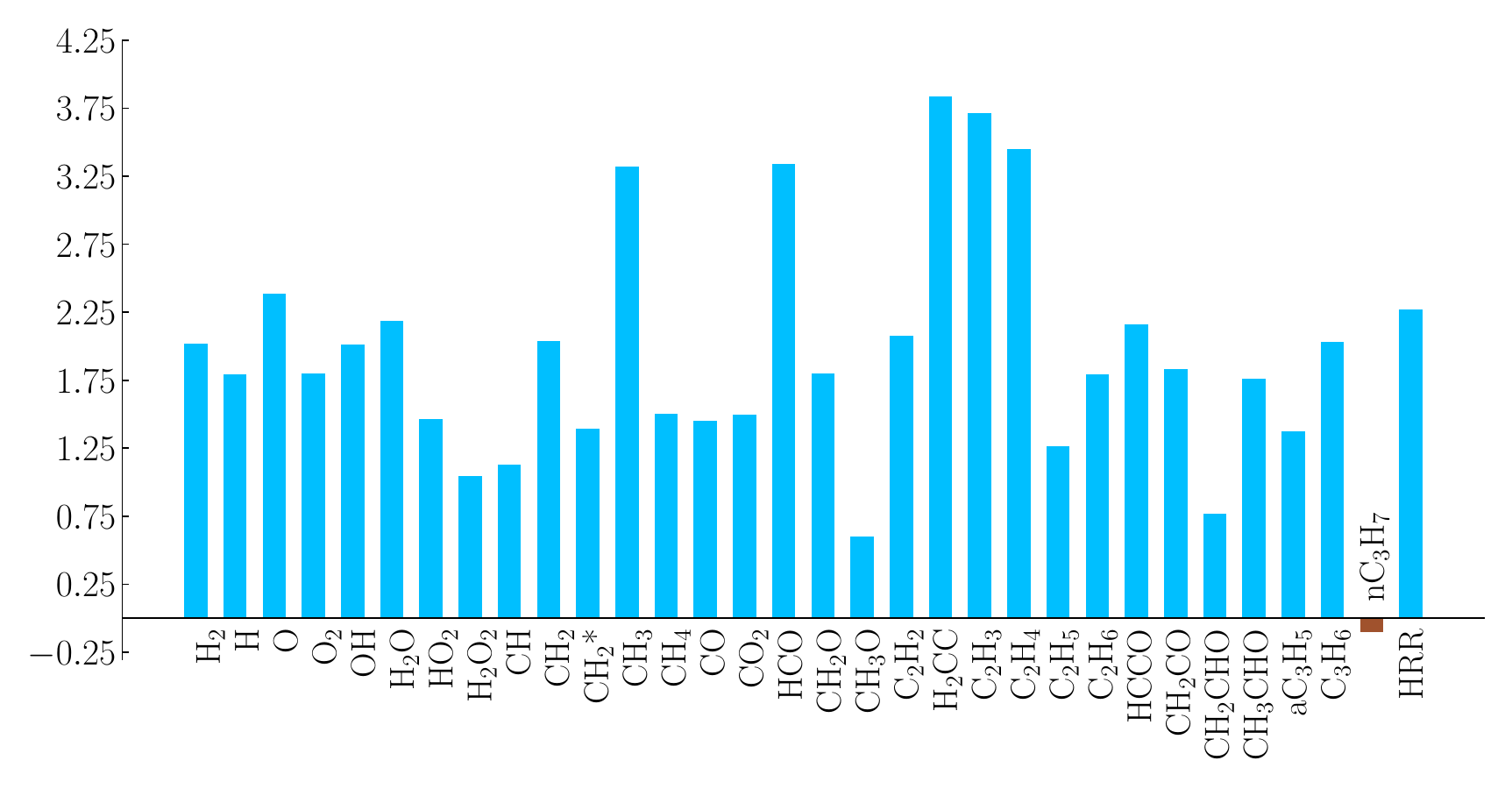}
    \begin{picture}(0,0)
        \put(-206,108){\scriptsize(a)}
        \put(-237,50){\scriptsize {\rotatebox{90}{Error ratio $r_a$}}}
    \end{picture}
    }\hspace{-0.018cm}
    {
    \includegraphics[width=7.98cm]{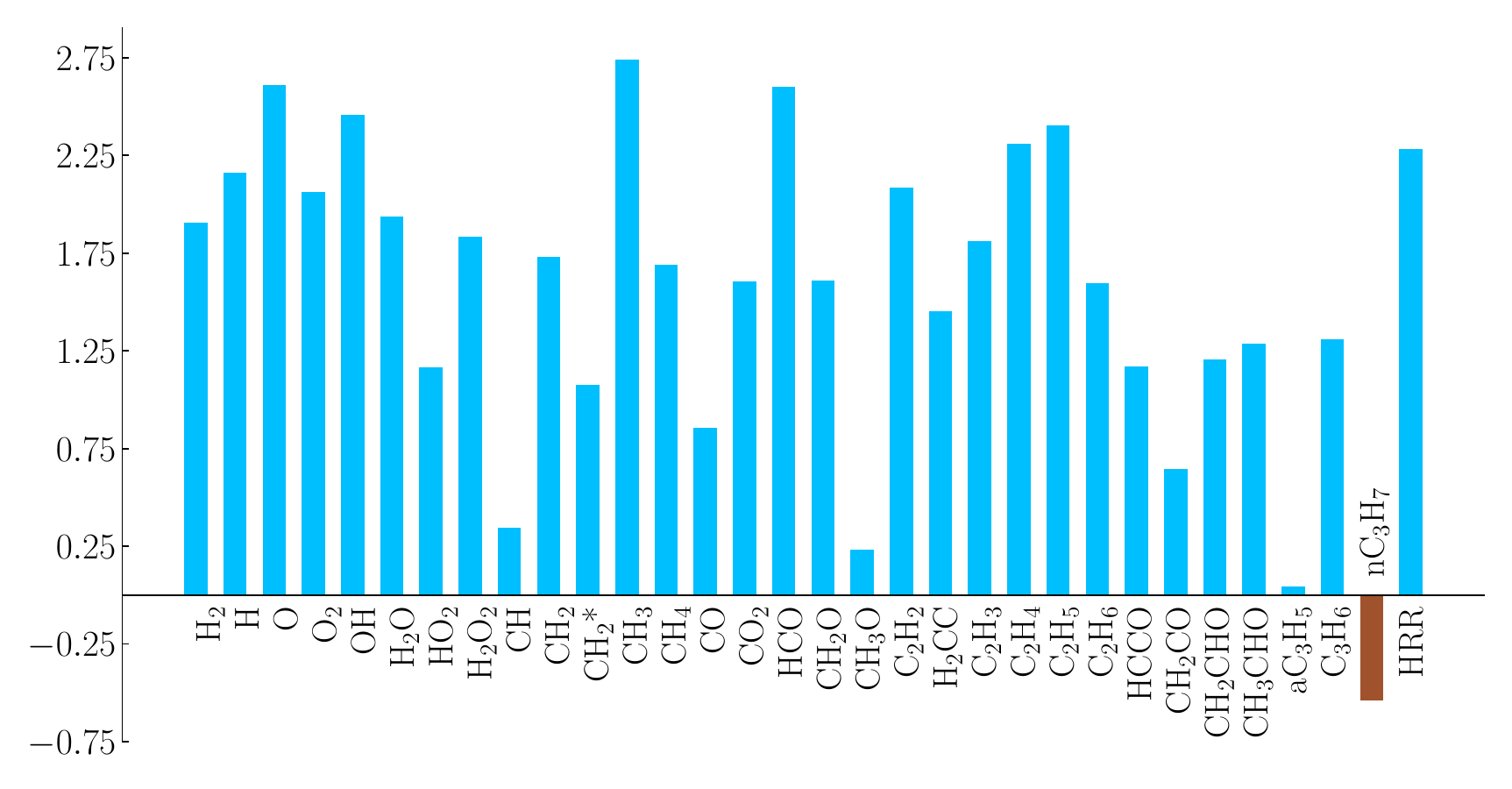}
    \begin{picture}(0,0)
        \put(-206,108){\scriptsize(c)}
        \put(-237,50){\scriptsize {\rotatebox{90}{Error ratio $r_a$}}}
    \end{picture}
    }
    {
    \includegraphics[width=7.98cm]{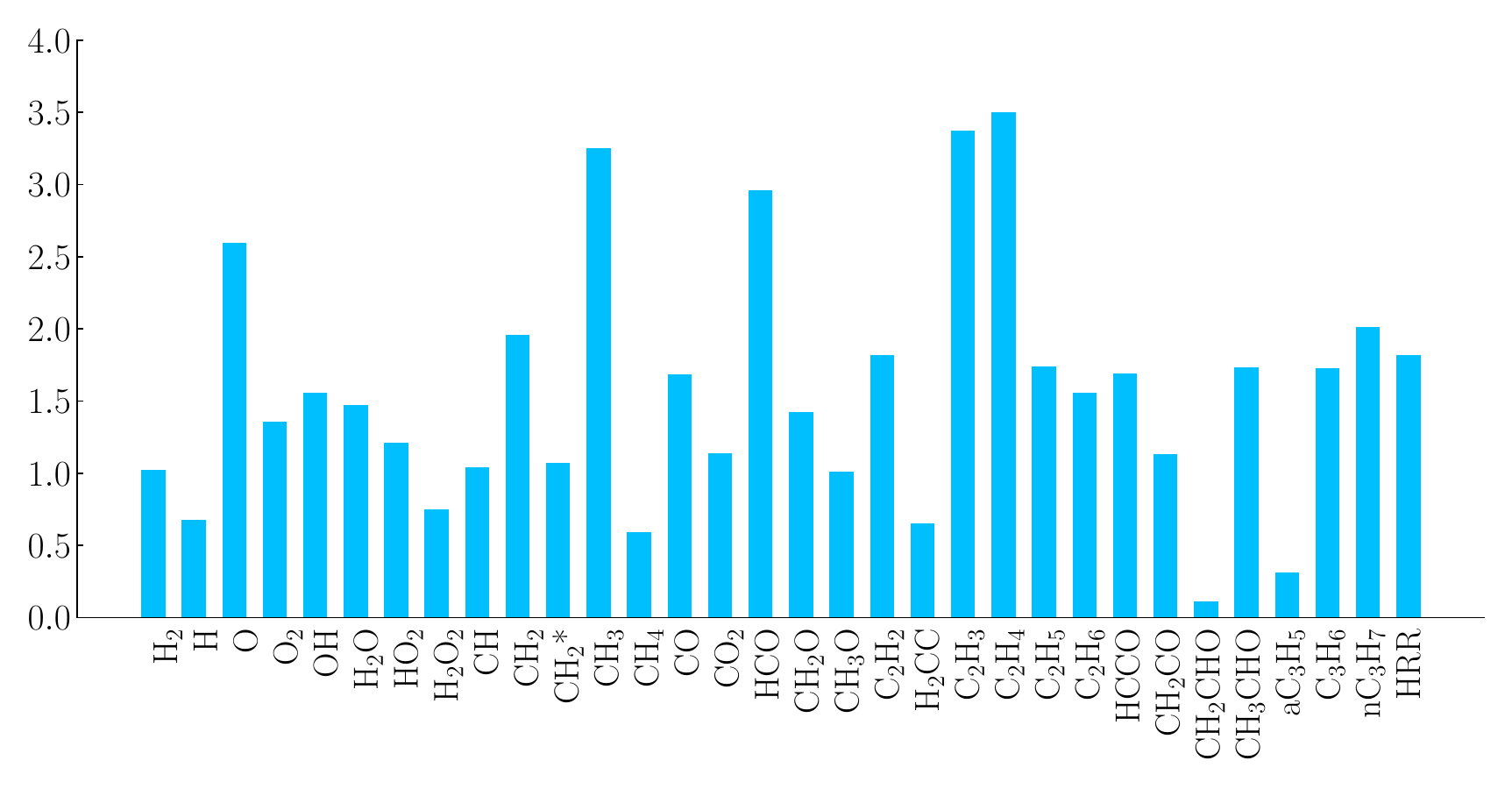}
    \begin{picture}(0,0)
        \put(-209,108){\scriptsize(b)}
        \put(-236,50){\scriptsize {\rotatebox{90}{Error ratio $r_m$}}}
    \end{picture}
    }\hspace{-0.018cm}
    {
    \includegraphics[width=7.98cm]{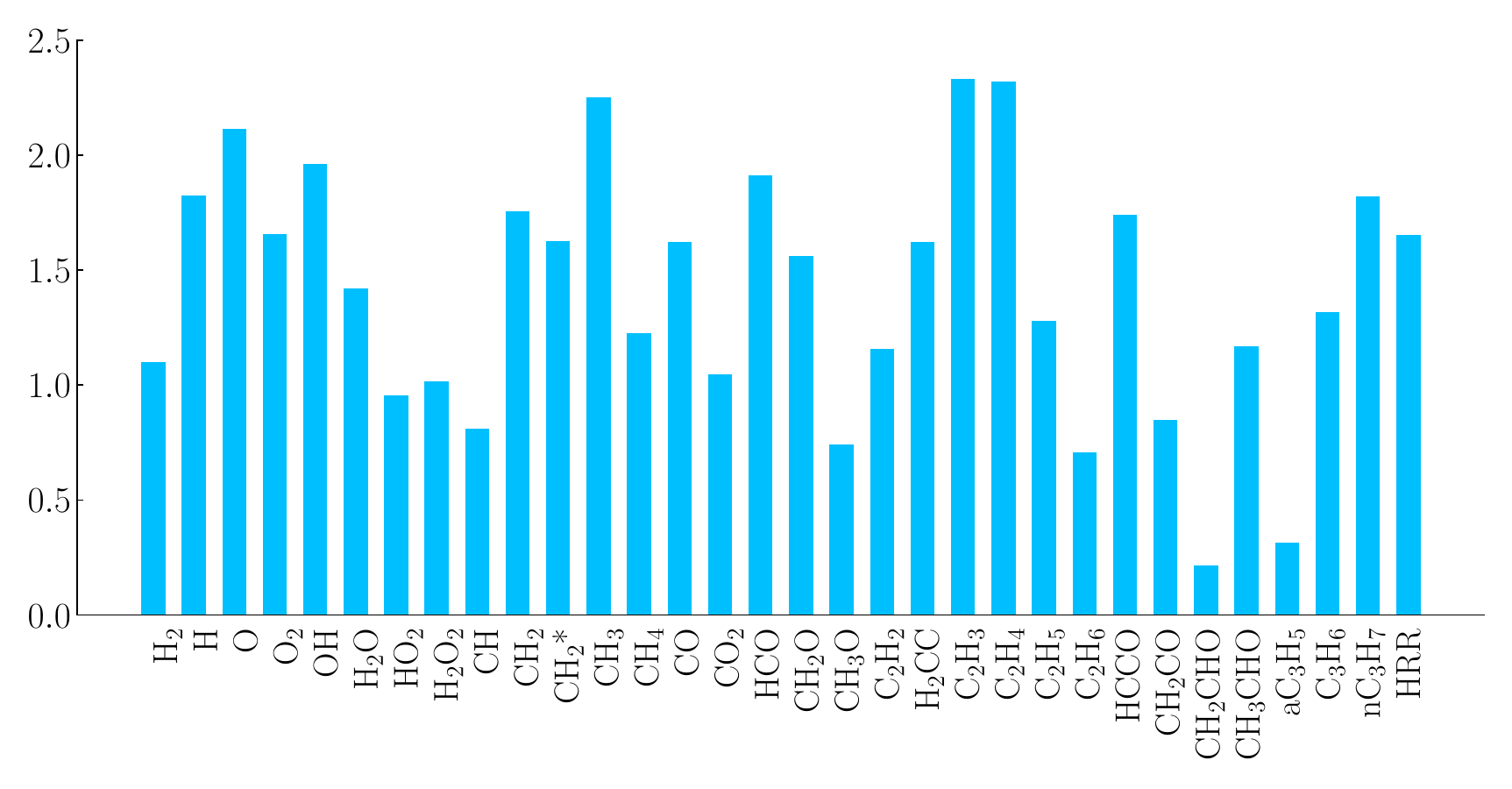}
    \begin{picture}(0,0)
        \put(-209,108){\scriptsize(d)}
        \put(-236,50){\scriptsize {\rotatebox{90}{Error ratio $r_m$}}}
    \end{picture}
    }
    \caption{Comparison of errors in the reconstruction of species production rates and heat release rate for (a), (b) CoK-PCA vs. CoK-PCA-ANN and (c), (d) PCA vs. PCA-ANN for the premixed ethylene-air homogeneous reactor dataset. Top and bottom plots in each column represent $r_a$ and $r_m$ respectively.}
    \label{fig:0D_9f_prs_linear_nonlinear}
\end{figure*}
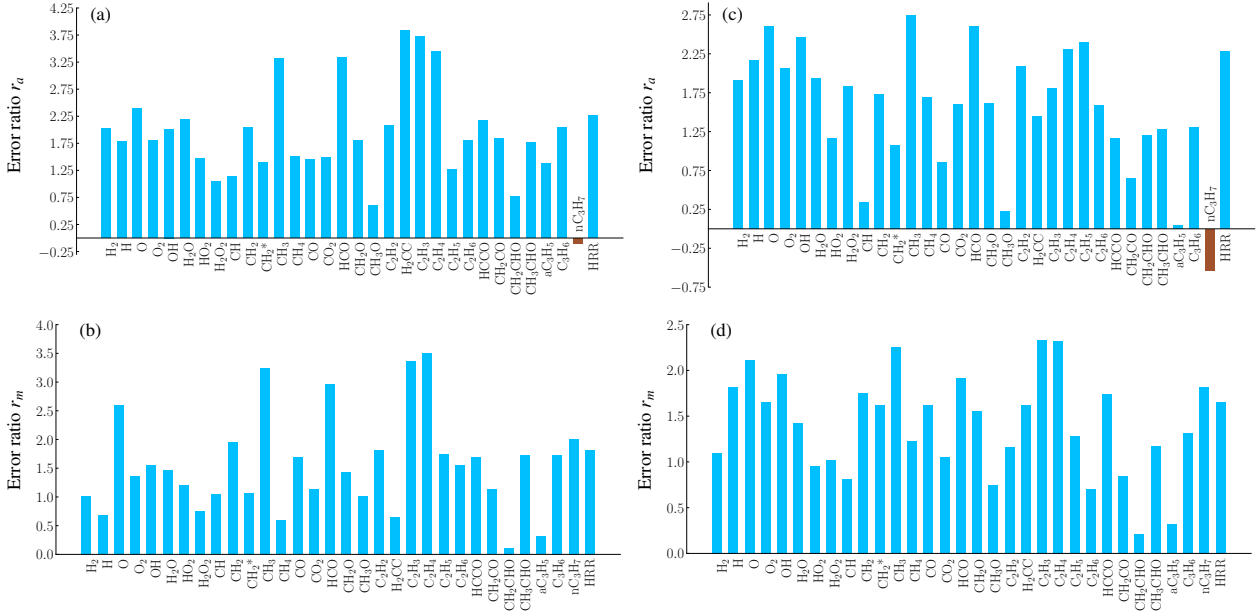

In Fig.~\ref{fig:0D_9f_tsc_linear_nonlinear}, we compare error ratios of linear and ANN reconstruction (Eqs.~\ref{eq:error-ratio2} and \ref{eq:error-ratio3}) of thermo-chemical scalars for both the dimensionality reduction methods. $\mathrm{N_2}$ being an inert species has not been included here. For most variables, ANN reconstruction outperforms linear reconstruction (demonstrated by blue bars) with respect to the average ($r_a$) and maximum ($r_m$) error metrics. An exception is temperature, where linear reconstruction performs marginally better in terms of $r_m$ (demonstrated by brown bars). This observation is consistent for both methods, i.e., PCA and CoK-PCA. 
Not surprisingly, as shown in Fig.~\ref{fig:0D_9f_prs_linear_nonlinear}, the errors in species production rates and heat release rate, computed from the reconstructed thermo-chemical state, are significantly lower with ANN reconstruction compared with linear reconstruction.  
In general, as $n_q$ increases, the accuracy improvements obtained with ANN in comparison to linear reconstruction decrease as the reduced manifold becomes an increasingly better linear approximation of the original state; in the limit of $n_q = n_v$ linear reconstruction is exact, which is a scenario with no reduction in dimensionality. As dimensionality needs to be reduced as aggressively as possible, one can conclude that ANN is better suited for reconstructing data from low-dimensional manifolds.
\begin{figure*}[h!]
    \centering
    {
    \includegraphics[width=7.98cm]{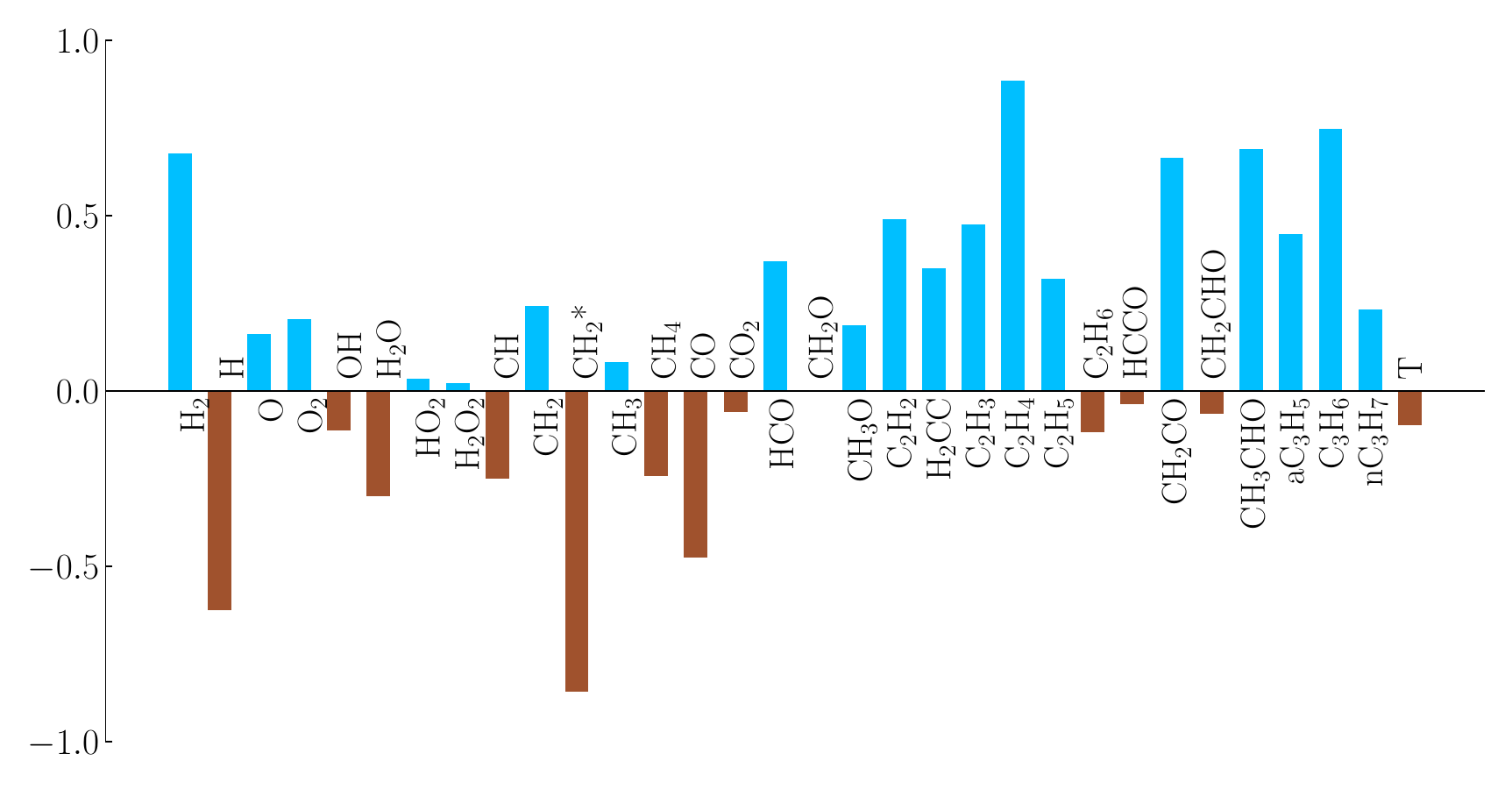}
    \begin{picture}(0,0)
        \put(-206,108){\scriptsize(a) Species mass fractions and temperature}
        \put(-237,50){\scriptsize {\rotatebox{90}{Error ratio $r_a$}}}
    \end{picture}
    }\hspace{-0.018cm}
    {
    \includegraphics[width=7.98cm]{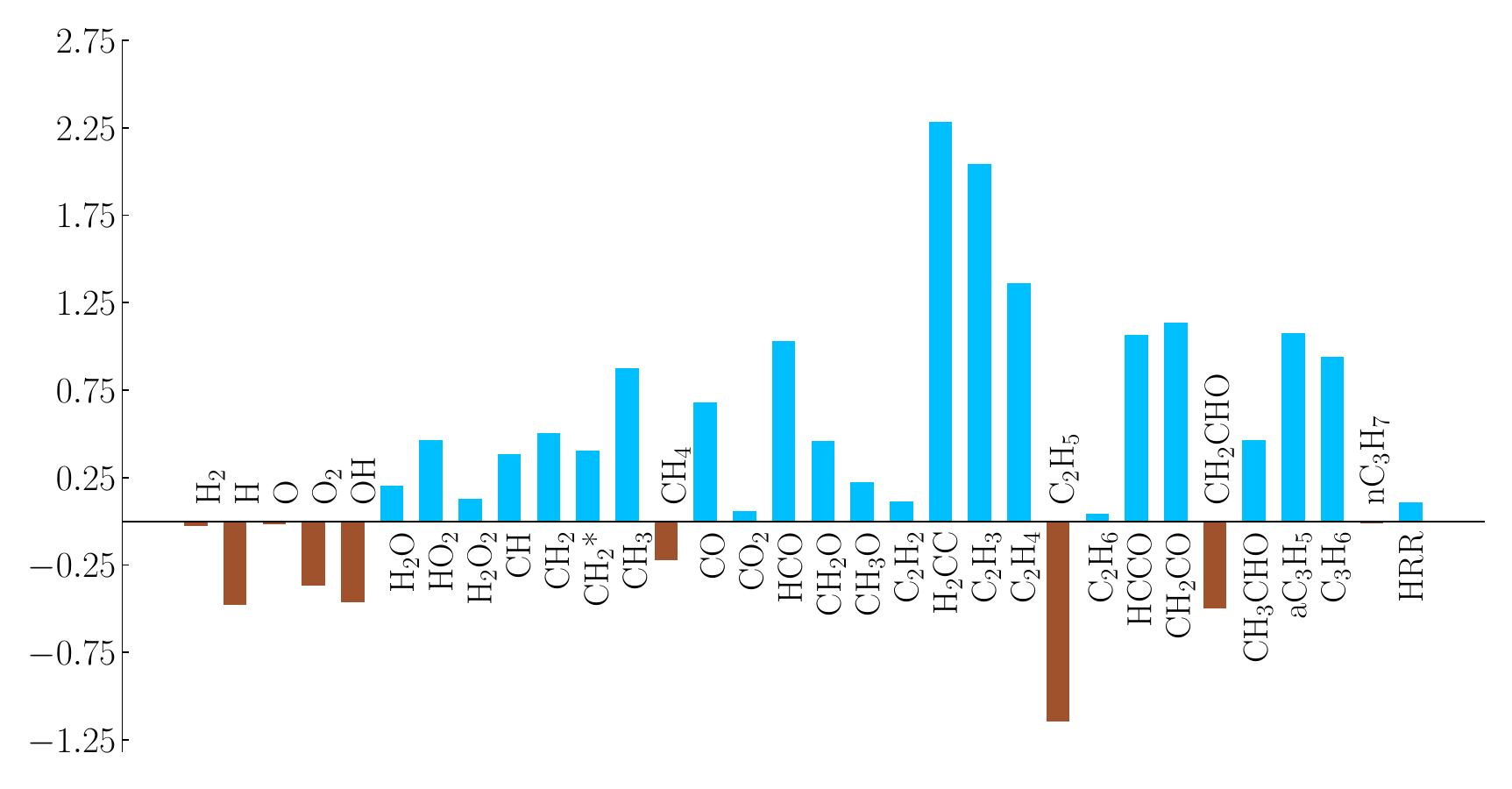}
    \begin{picture}(0,0)
        \put(-206,108){\scriptsize(c) Species production rates and heat release rate}
        \put(-237,50){\scriptsize {\rotatebox{90}{Error ratio $r_a$}}}
    \end{picture}
    }
    {
    \includegraphics[width=7.98cm]{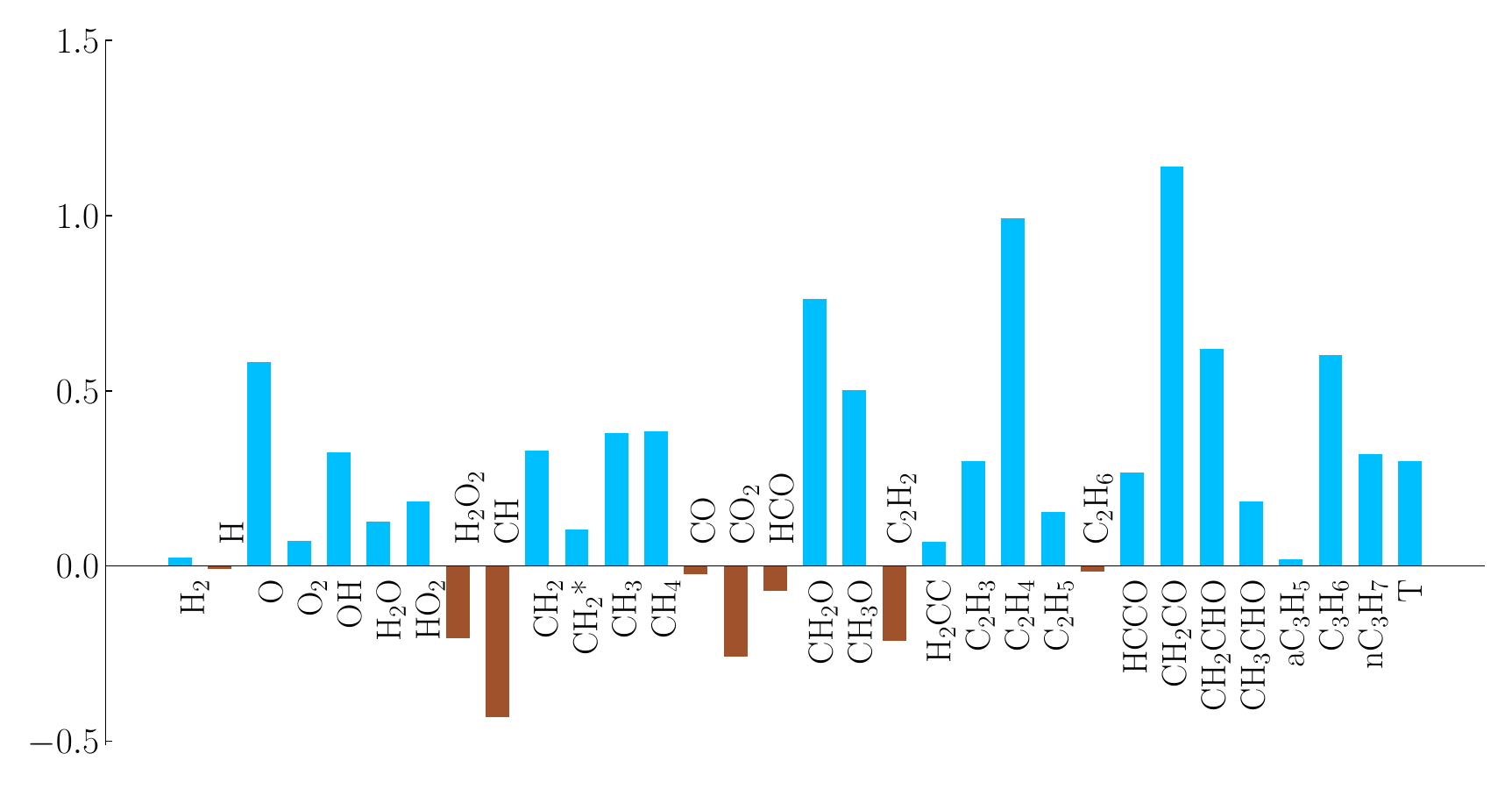}
    \begin{picture}(0,0)
        \put(-206,108){\scriptsize(b) Species mass fractions and temperature}
        \put(-236,50){\scriptsize {\rotatebox{90}{Error ratio $r_m$}}}
    \end{picture}
    }\hspace{-0.018cm}
    {
    \includegraphics[width=7.98cm]{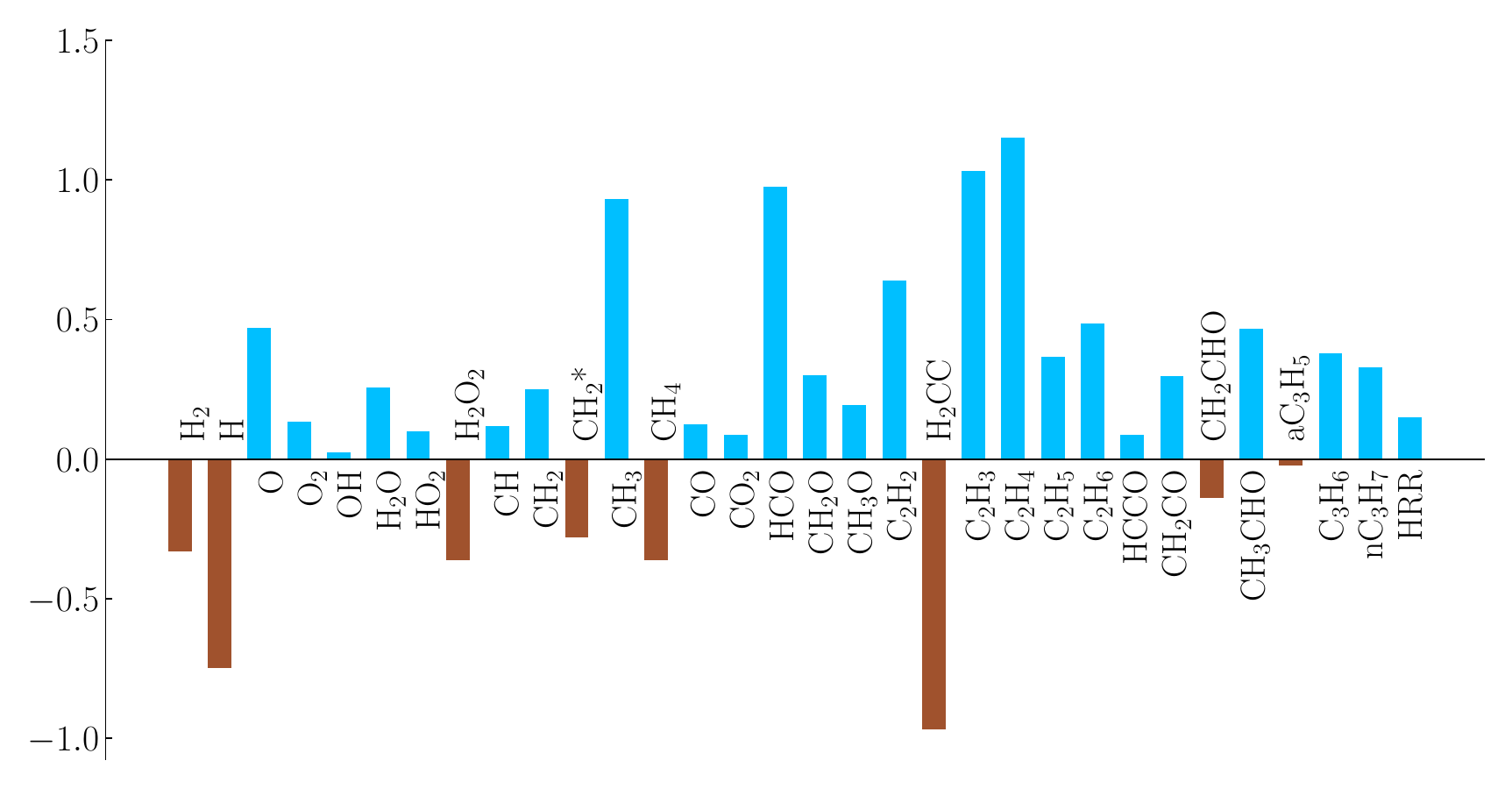}
    \begin{picture}(0,0)
        \put(-206,108){\scriptsize(d) Species production rates and heat release rate}
        \put(-236,50){\scriptsize {\rotatebox{90}{Error ratio $r_m$}}}
    \end{picture}
    }
    \caption{Comparison of errors in the reconstruction of thermo-chemical scalars (left), species production rates and heat release rate (right) for PCA-ANN vs. CoK-PCA-ANN for the premixed ethylene-air homogeneous reactor dataset. Top and bottom plots in each column represent $r_a$ and $r_m$ respectively.}
    \label{fig:0D_9f_nonlinear_pca_cok-pca}
\end{figure*}

Next, we compare the two dimensionality reduction techniques against each other, both with ANN reconstruction. Figure~\ref{fig:0D_9f_nonlinear_pca_cok-pca} shows the error ratios for PCA-ANN vs. CoK-PCA-ANN (Eq.~\ref{eq:error-ratio4}) in reconstructing thermo-chemical scalars (left), and species production rates and heat release rates (right). For the scalars, it can be clearly seen that CoK-PCA-ANN outperforms PCA-ANN in predictions of 25 and 21 (out of 33) variables for $r_m$ and $r_a$ metrics, respectively. The trend becomes more prominent in the case of species production rates and heat release rates where CoK-PCA-ANN predicts production rates more accurately for 23 out of the 32 species with the $r_a$ metric and 24 out of the 32 species with the $r_m$ metric. Notably, CoK-PCA-ANN captures heat release rate better than PCA-ANN in terms of both the error metrics. 
\begin{figure*}[h!]
    \centering
    {
    \includegraphics[width=6.5cm]{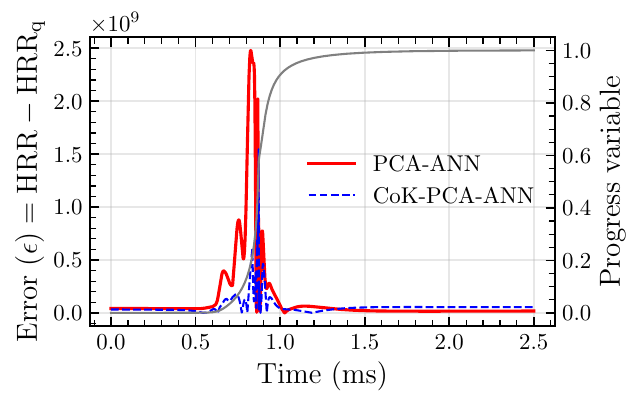}
    \begin{picture}(0,0)
        \put(-153,95){\scriptsize(a)}
    \end{picture}
    }
    \hspace{0.2cm}
    {
    \includegraphics[width=6.5cm]{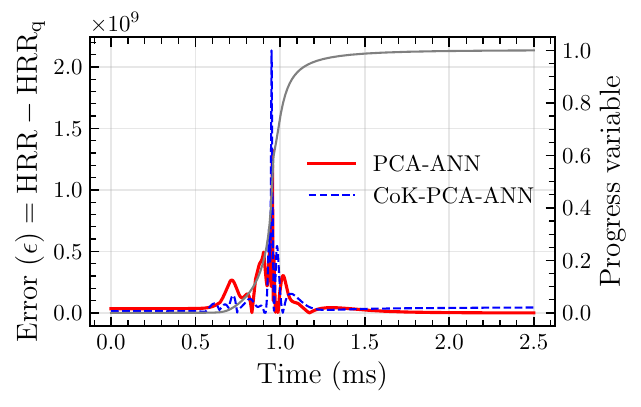}
    \begin{picture}(0,0)
        \put(-153,95){\scriptsize(c)}
    \end{picture}
    }
    \vspace{-0.35cm}
    {
    \includegraphics[width=6.5cm]{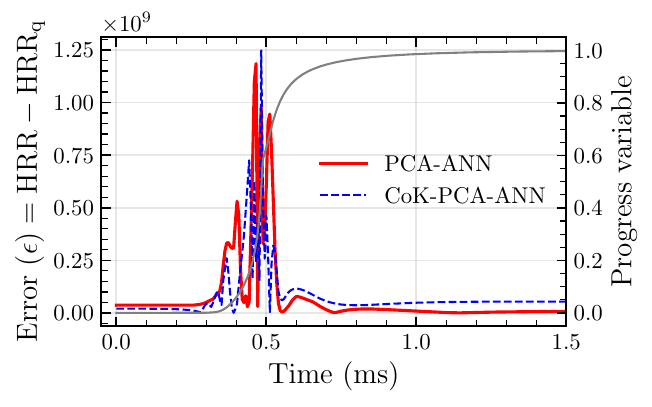}
    \begin{picture}(0,0)
        \put(-152,93){\scriptsize(b)}
    \end{picture}
    }
    \hspace{0.2cm}
    {
    \includegraphics[width=6.5cm]{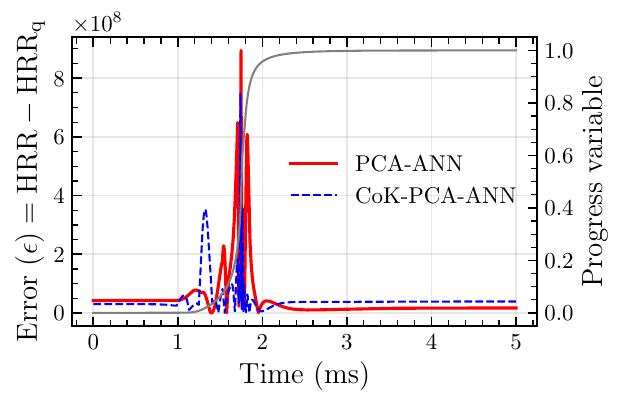}
    \begin{picture}(0,0)
        \put(-157,95){\scriptsize(d)}
    \end{picture}
    }
    \caption{Temporal evolution of absolute errors in reconstructed heat release rates for the test states - (a) $D_3$, (b) $D_5$, (c) $D_7$, and (d) $D_9$ for the premixed ethylene-air homogeneous reactor dataset. The progress variable is plotted in grey for reference.}
    \label{fig:0D_9f_cok-ann-pca-ann_hrr}
\end{figure*}

While $r_m$ and $r_a$ are global error metrics, it is instructive to examine the temporal distribution of reconstruction errors and determine whether the errors are low/high in the unburnt, igniting, or fully burnt portions of the flame. Figure~\ref{fig:0D_9f_cok-ann-pca-ann_hrr} presents the absolute reconstruction error of heat release rate plotted against time for the four test flamelets: $D_3, D_5, D_7,$ and $D_9$. For reference, the progress variable is plotted on the right $y$-axis of each figure. Both methods incur significant error in the reaction zones, with the peak at intermediate values of the progress variable, which occurs at \SI{0.8}{\milli\second}, \SI{0.4}{\milli\second}, \SI{1}{\milli\second}, and \SI{1.9}{\milli\second} for $D_3, D_5, D_7,$ and $D_9$, respectively.
As expected, the error is much lower on the unburnt and the fully burnt portions. Further, for $D_3$ and $D_9$, CoK-PCA-ANN incurs a significantly lower peak reconstruction error than PCA-ANN (demonstrated by the blue peaks smaller in magnitude than the red peaks), which is reflected in the $r_m$ error presented in Fig.~\ref{fig:0D_9f_nonlinear_pca_cok-pca} (d). However, the peak error for $D_7$ is higher for CoK-PCA-ANN. For $D_5$, both the methods incur essentially the same magnitude of errors and perform at par with each other. 
Nonetheless, across the four test flamelets, CoK-PCA-ANN yields an overall smaller average reconstruction error than PCA-ANN, as reflected in the $r_a$ error presented in Fig.~\ref{fig:0D_9f_nonlinear_pca_cok-pca} (c). These comparisons provide further evidence that the proposed CoK-PCA-ANN method predicts the overall chemical kinetics in the reaction zone better than PCA-ANN.

\subsection{Two-stage autoignition of dimethyl ether-air mixture}\label{sec:0D_2stage_DME}
In contrast to ethylene, which has conventional single-stage ignition chemistry, a class of hydrocarbon fuels characterized by more complex two-stage ignition (a low-temperature and a high-temperature) chemistry are increasingly considered suitable for novel combustion concepts such as homogeneous charge compression ignition (HCCI)~\cite{BansalMC2015}. HCCI relies on volumetric autoignition of a (nearly) homogeneous fuel charge and realizes the benefits of low emissions due to fuel-lean combustion while also achieving high efficiencies. However, controlling the ignition timing is the biggest challenge since the charge ignites spontaneously due to compression heating. Consequently, modeling the ignition processes of two-stage ignition fuels under engine-relevant conditions is an open challenge. Dimethyl ether (DME) is a prominent example, and its ignition behavior resulting from turbulence-chemistry interactions at engine-relevant conditions has been widely studied using DNS~\cite{BansalMC2015, BhagatwalaLSSLC2015, KrismanHTBC2017}. From a dimensionality reduction perspective, DME ignition presents distinct challenges from that of ethylene; the chemical pathways and the participating chemical species for the low-temperature ignition chemistry are different from high-temperature chemistry. This motivates us to test the capability of CoK-PCA-ANN in reconstructing the original state space from the reduced manifold for the two-stage ignition of DME.

We consider a constant pressure zero-dimensional homogeneous reactor of a stoichiometric mixture of hydrogen-enriched DME fuel and air. The ratio of hydrogen to DME is 3:2 in the fuel mixture, similar to that in~\cite{BhagatwalaLSSLC2015}. The initial pressure is 1 atm while the initial temperature is varied from \SI{600}{K} to \SI{800}{K} in increments of \SI{25}{K}, for a total of nine flames. This range of initial temperatures is such that the flames contain both two-stage as well as single-stage ignition behavior. Finite rate chemistry is specified using the 39-species, 175-reactions skeletal mechanism developed in~\cite{BhagatwalaLSSLC2015}, and the flames are simulated with Cantera~\cite{Cantera} for a duration of \SI{1}{\second} with a fixed time step of \SI{0.1}{\milli\second}. In this case, the original design matrix $\mathbf{D}$ consists of $n_g = 10001$ points and $n_v = 40$ variables, comprising 39 species and temperature.

Traditional dimensionality reduction techniques, such as PCA or linear regression, may not effectively capture the nonlinear interactions present in the data. The data associated with the two-stage ignition of DME is high-dimensional and contains intricate patterns. This includes time-dependent or transient behavior, multiple ignition modes, and variations under different operating conditions. This complexity makes it difficult to find a low-dimensional representation that captures the essential information while discarding irrelevant or redundant features. The reconstruction of two-stage ignition using CoK-PCA-ANN offers several benefits. It enables a deeper understanding of DME combustion, facilitates the development of more accurate ignition models, and provides valuable insights for optimizing combustion strategies. This approach aids in reducing data dimensionality by extracting pertinent features and eliminating redundant information, thereby improving computational efficiency while maintaining prediction accuracy.

CoK-PCA and PCA are performed using the data of all nine flames, and dimensionality is reduced to $n_q=5$. To train the ANNs for reconstructing the full thermo-chemical state from the reduced state, the data is split into training and testing sets, with five flames (initial temperatures of 600 K, 650 K, 700 K, 750 K, 800 K) comprising the former, and the rest, the latter. We randomly shuffle the training dataset and set aside $20 \%$ for the validation process. After conducting hyperparameter tuning, the network architecture is determined with two hidden layers comprising 10 and 20 neurons, respectively. The input and output layers have a width of $n_q = 5$ and $n_v = 40$ neurons, respectively. A hyperbolic tangent activation function for the hidden layers, a custom sigmoid-based activation function for the output layer, and the Adam optimizer are used as before.

\begin{figure}[h!]
    \centering
    {
    \includegraphics[width = 6cm]{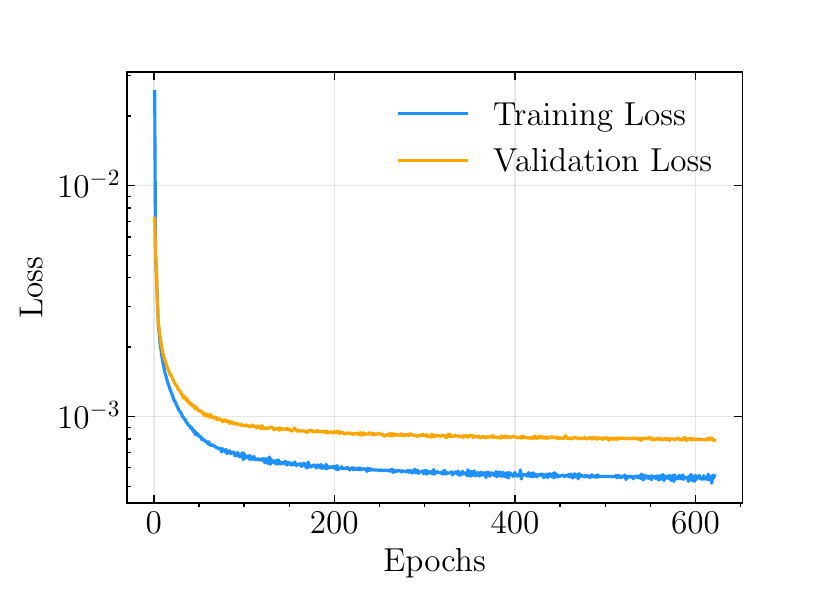}%
    \begin{picture}(0,0)
        \put(-130,90){\scriptsize (a)}
    \end{picture}
    }
    {
    \includegraphics[width = 6cm]{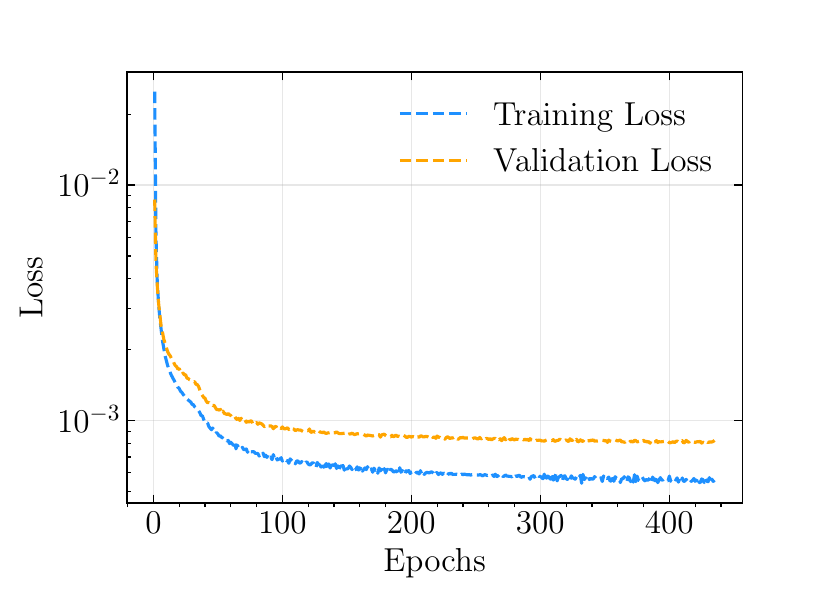}
    \begin{picture}(0,0)
        \put(-130,90){\scriptsize (b)}
    \end{picture}
    }
    \caption{Training and validation loss curves for (a) CoK-PCA-ANN and (b) PCA-ANN, respectively, for the DME two-stage autoignition dataset.}
    \label{fig:DME_train_validation_loss}
\end{figure}
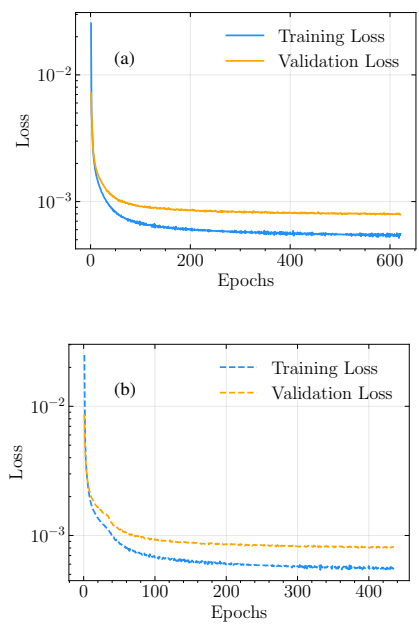

Figure~\ref{fig:DME_train_validation_loss} shows the training and validation loss for the PCA-ANN and CoK-PCA-ANN. It is evident that the validation loss remains consistently only slightly higher than the training loss ($\sim$ \num{2.5e-4}) for a significant number of epochs (200-500), and the model has converged. We employ early stopping to achieve this convergence, thereby saving computational resources and preventing overfitting. This indicates that the model is generalizing well to unseen data. Despite the slight difference in loss, the model demonstrates robustness and reliability in its predictions. This suggests that the model has learned intricate patterns present in the two-stage ignition dataset and features from the training data that allow it to make accurate predictions on new examples, resulting in a reliable and effective model.
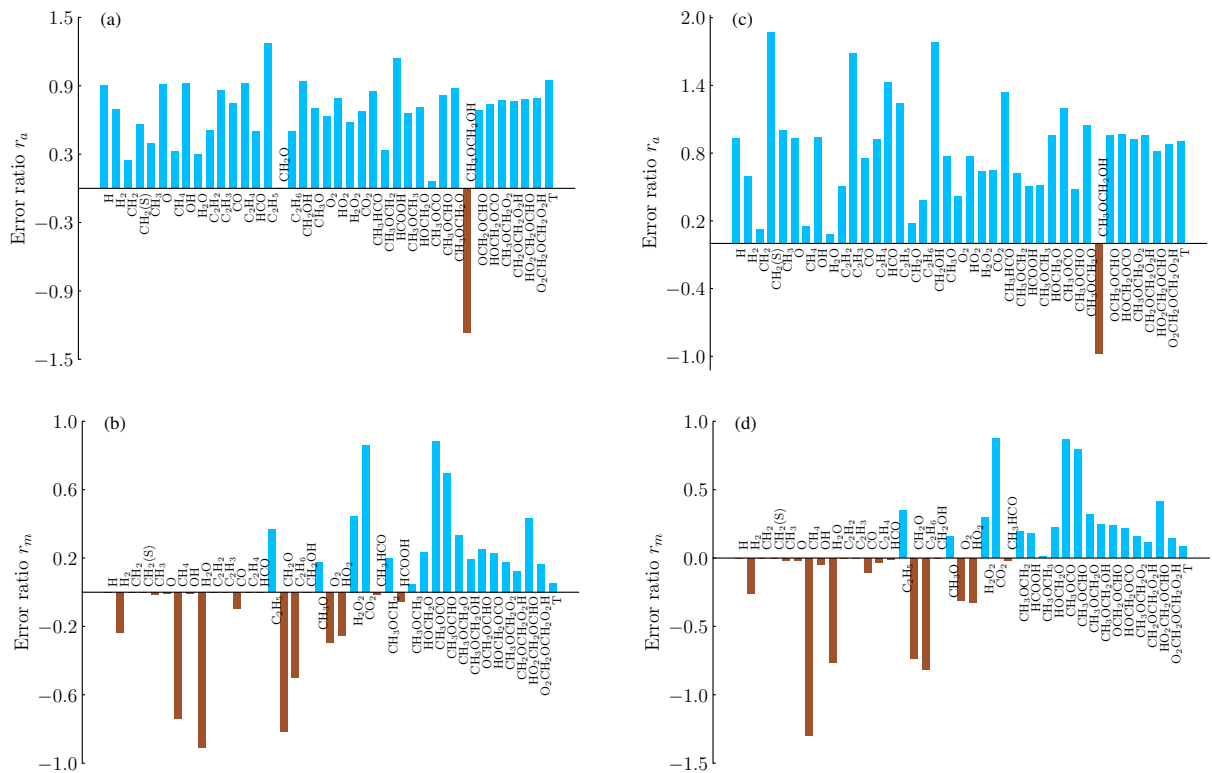
\begin{figure*}[h!]
    \centering
    {
    \includegraphics[width = 7.98cm]{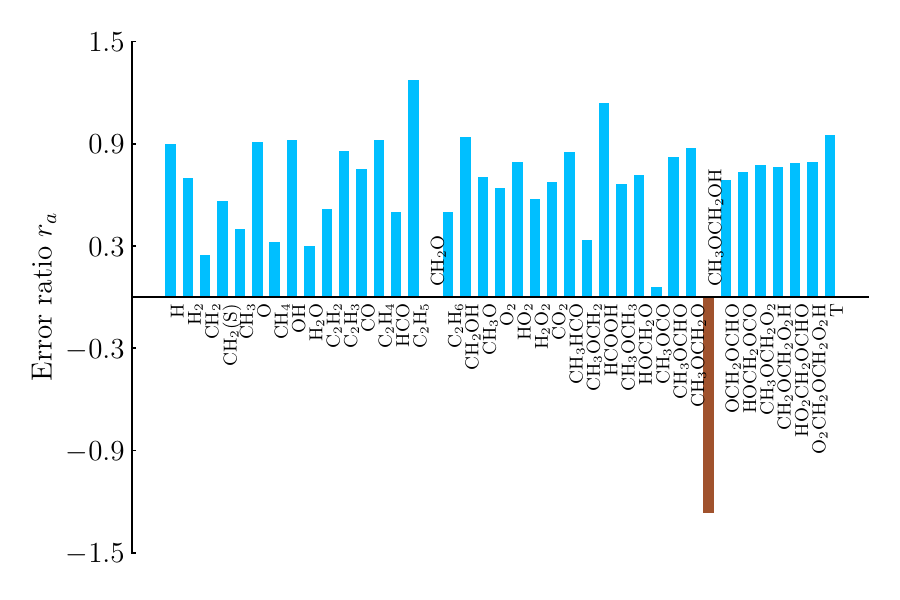}
    \begin{picture}(0,0)
        \put(-188,138){\scriptsize(a)}
    \end{picture}
    }\hspace{-0.018cm}
    {
    \includegraphics[width = 7.98cm]{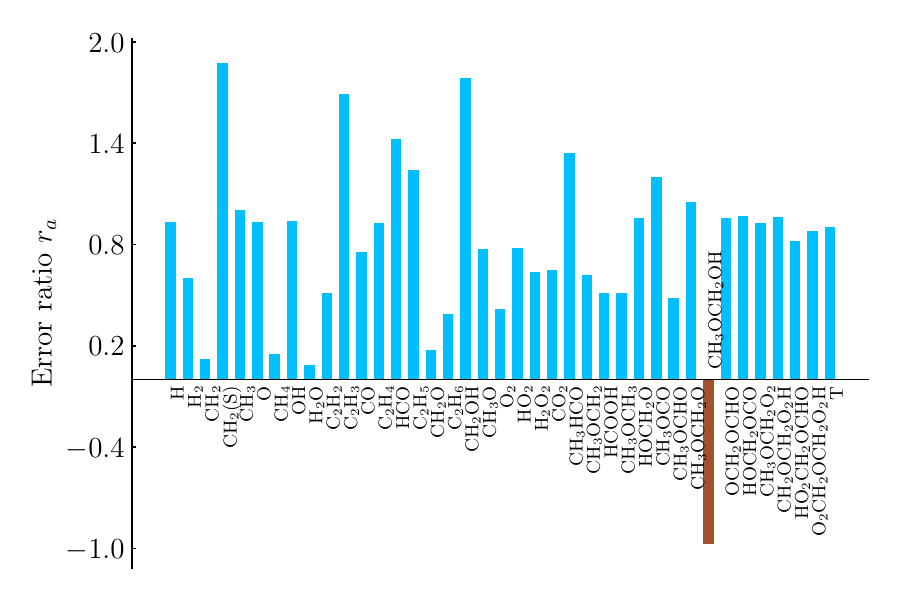}
    \begin{picture}(0,0)
        \put(-188,138){\scriptsize(c)}
    \end{picture}
    }
\vspace{-0.35cm}    
    {
    \includegraphics[width = 7.98cm]{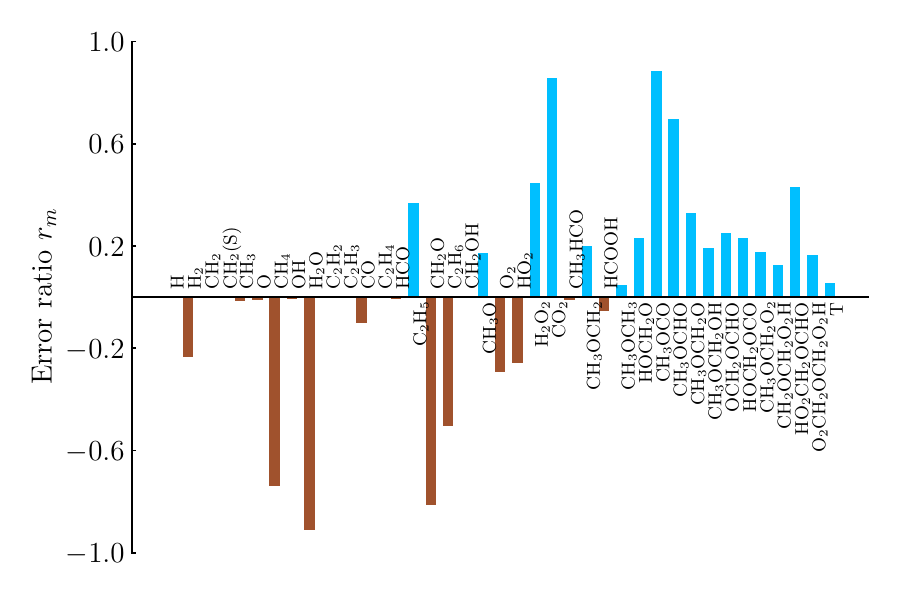}
    \begin{picture}(0,0)
        \put(-188,138){\scriptsize(b)}
    \end{picture}
    }\hspace{-0.018cm}
    {
    \includegraphics[width = 7.98cm]{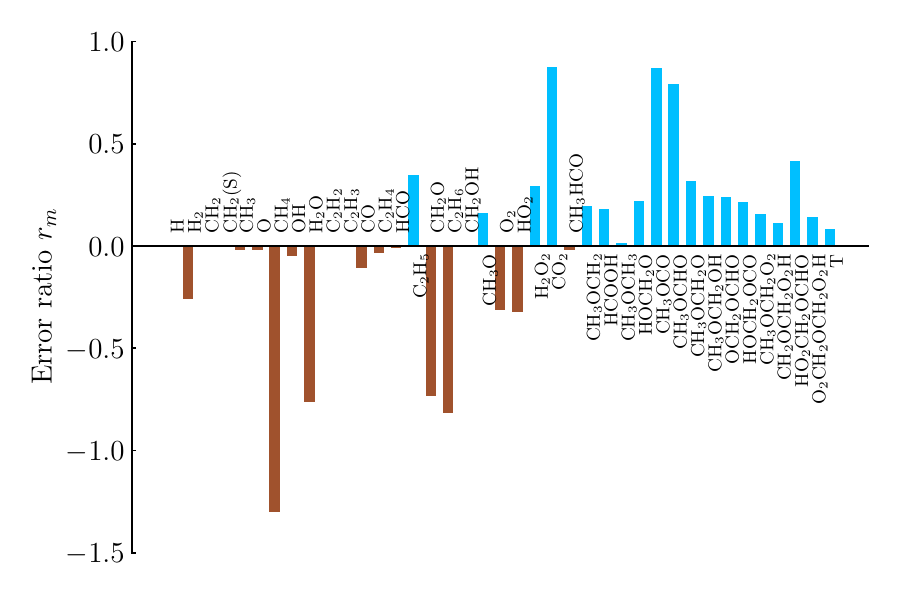}
    \begin{picture}(0,0)
        \put(-188,138){\scriptsize(d)}
    \end{picture}
    }
\vspace{-0.45cm}     
    \caption{Comparison of errors in the reconstruction of thermo-chemical scalars of the DME two-stage autoignition dataset for (a), (b) CoK-PCA vs. CoK-PCA-ANN and (c), (d) PCA vs. PCA-ANN. Top and bottom plots in each column represent $r_a$ and $r_m$ respectively.}
    \label{fig:DME_scalar_recon_errors}
\end{figure*}

\begin{figure*}[h!]
    \centering
    {
    \includegraphics[width = 7.98cm]{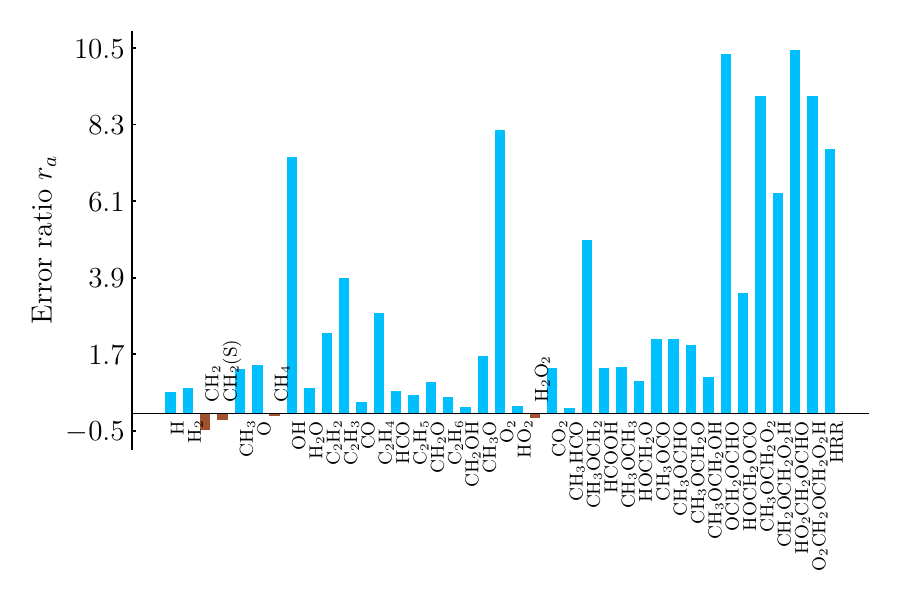}
    \begin{picture}(0,0)
        \put(-188,138){\scriptsize(a)}
    \end{picture}
    }\hspace{-0.018cm}
    {
    \includegraphics[width = 7.98cm]{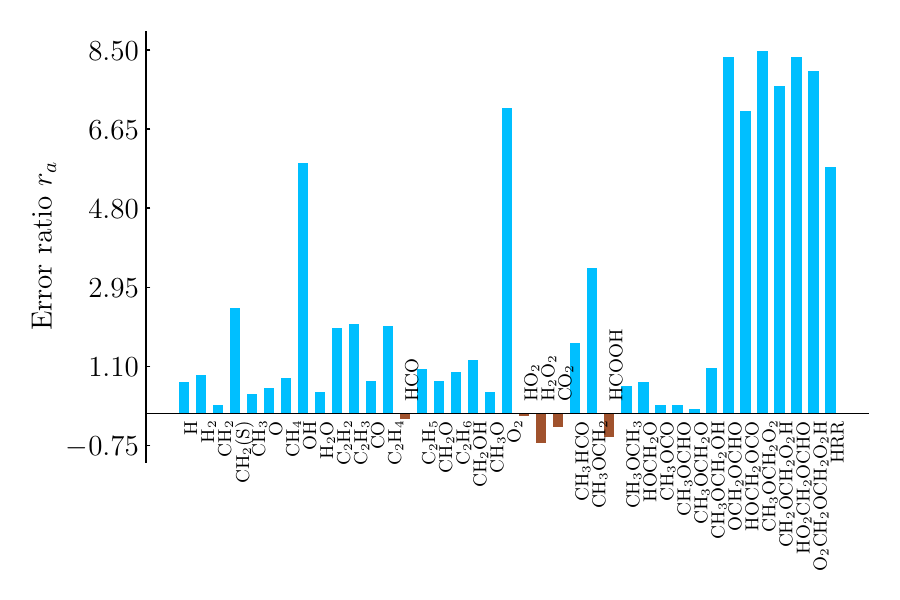}
    \begin{picture}(0,0)
        \put(-188,138){\scriptsize(c)}
    \end{picture}
    }
    \vspace{-0.30cm}
    {
    \includegraphics[width = 7.98cm]{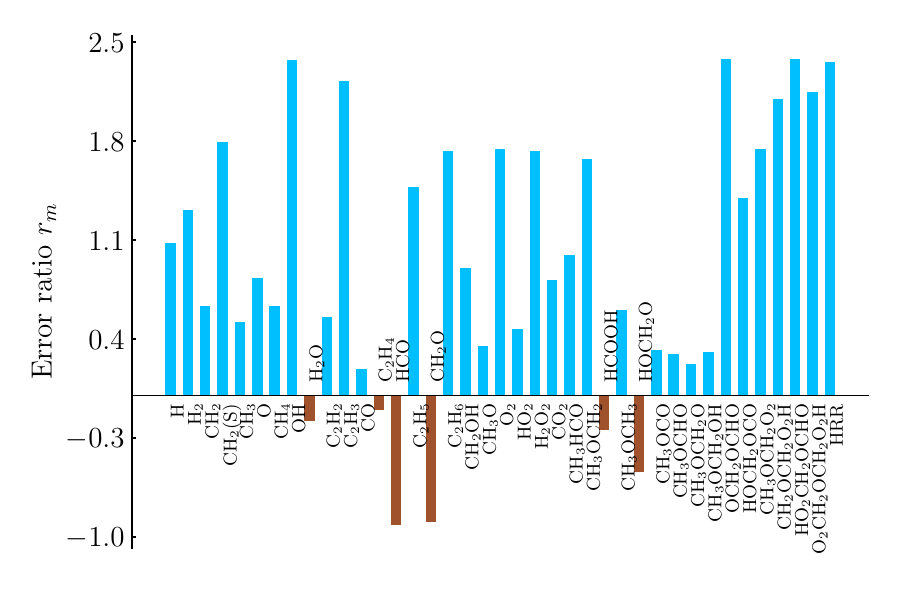}
    \begin{picture}(0,0)
        \put(-188,138){\scriptsize(b)}
    \end{picture}
    }\hspace{-0.018cm}
    {
    \includegraphics[width = 7.98cm]{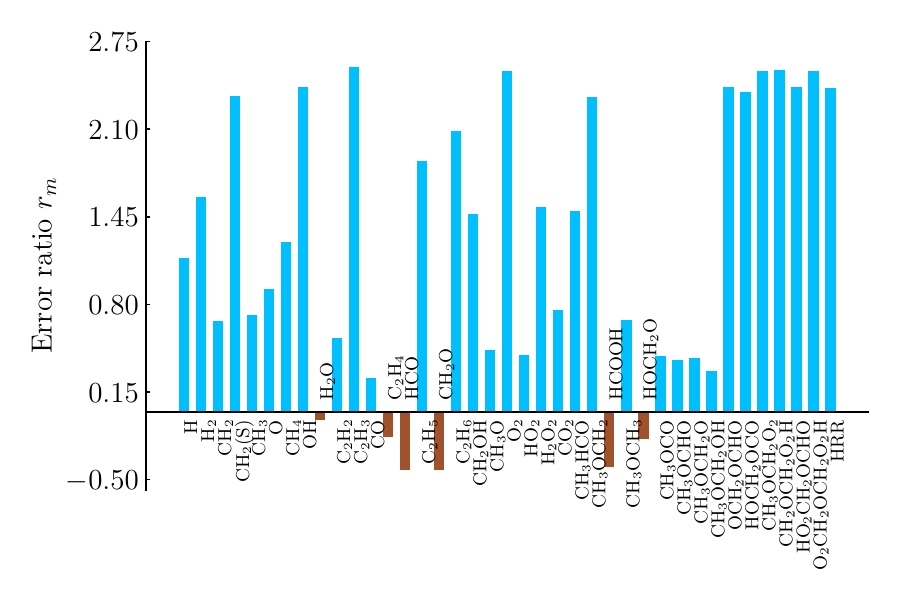}
    \begin{picture}(0,0)
        \put(-188,138){\scriptsize(d)}
    \end{picture}
    }
    \vspace{-0.45cm}
    \caption{Comparison of errors in the reconstruction of species production rates and heat release rate of the DME two-stage autoignition dataset for (a), (b) CoK-PCA vs. CoK-PCA-ANN and (c), (d) PCA vs. PCA-ANN. Top and bottom plots in each column represent $r_a$ and $r_m$ respectively.}
    \label{fig:DME_reac_rate_recon_errors}
\end{figure*}

\begin{figure*}[h!]
    \centering
    {
    \includegraphics[width = 7.98cm]{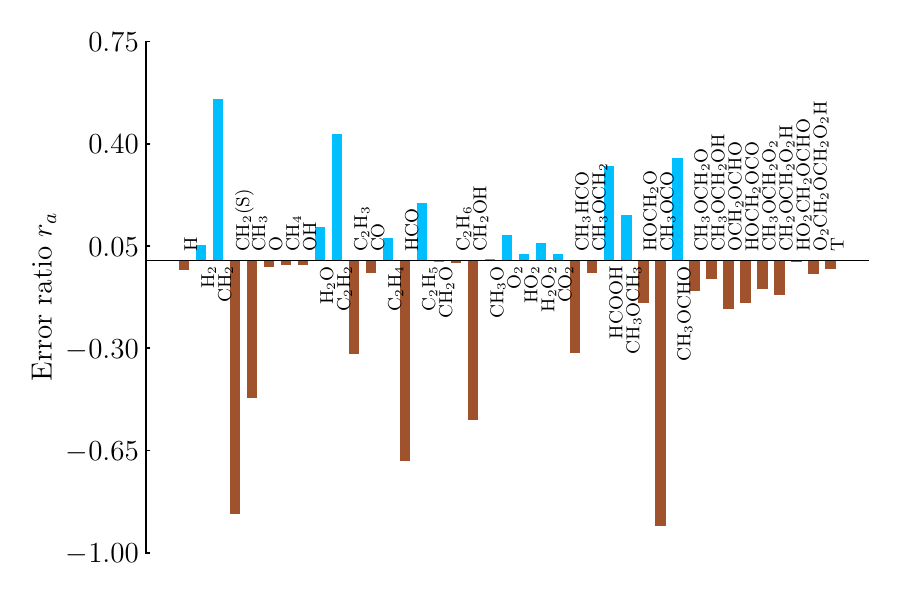}
    \begin{picture}(0,0)
        \put(-187,140){\scriptsize(a) Species mass fractions and temperature}
    \end{picture}
    }\hspace{-0.018cm}
    {
    \includegraphics[width = 7.98cm]{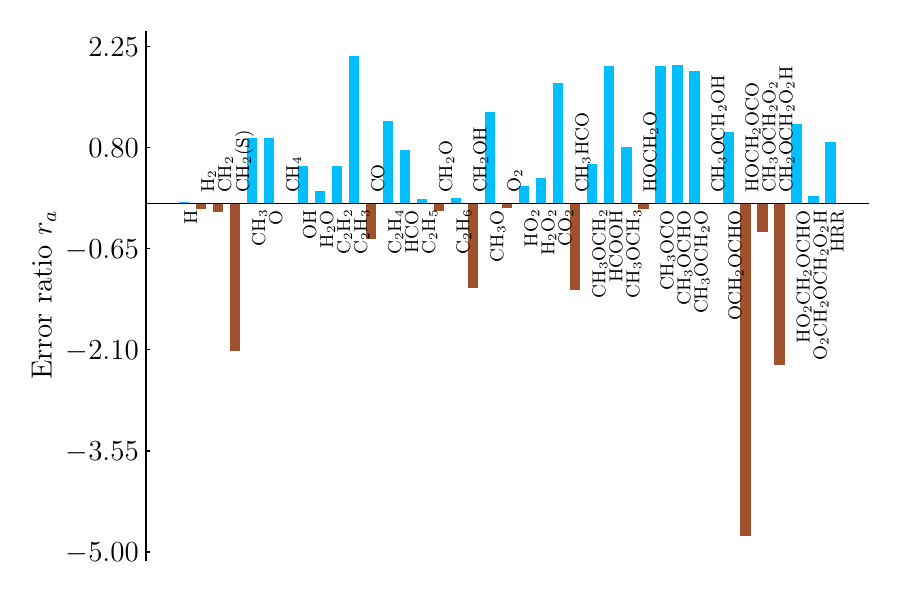}
    \begin{picture}(0,0)
        \put(-187,140){\scriptsize(c) Species production rates and heat release rate}
    \end{picture}
    }
    {
    \includegraphics[width = 7.98cm]{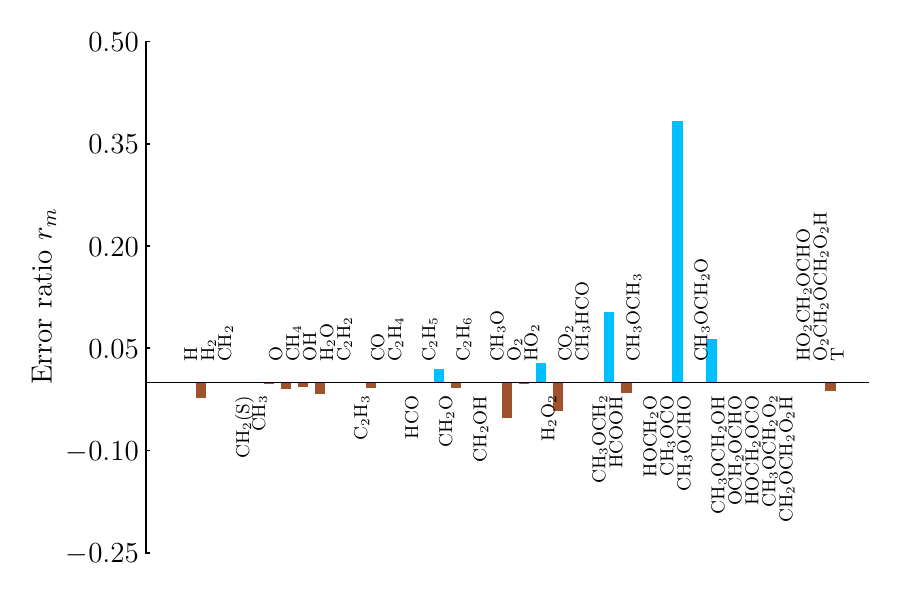}
    \begin{picture}(0,0)
        \put(-187,140){\scriptsize(b) Species mass fractions and temperature}
    \end{picture}
    }\hspace{-0.018cm}
    {
    \includegraphics[width = 7.98cm]{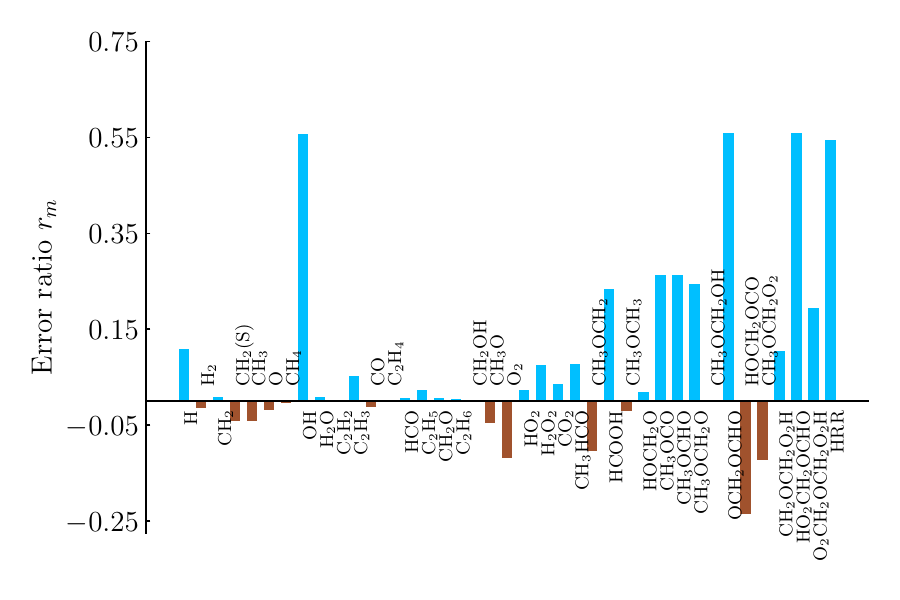}
    \begin{picture}(0,0)
        \put(-187,140){\scriptsize(d) Species production rates and heat release rate}
    \end{picture}
    }
    \vspace{-0.45cm}
    \caption{Comparison of errors in the reconstruction of thermo-chemical scalars (left), species production rates and heat release rate (right) for PCA-ANN vs. CoK-PCA-ANN of the DME two-stage autoignition dataset. Top and bottom plots in each column represent $r_a$ and $r_m$ respectively.}
    \label{fig:DME_scalar_reac_rate_recon_errors}
\end{figure*}

The error ratios in thermo-chemical scalars, species production rates, and heat release rates were computed using equations (\ref{eq:error-ratio2}) - (\ref{eq:error-ratio4}) and visualized in Figures \ref{fig:DME_scalar_recon_errors}, \ref{fig:DME_reac_rate_recon_errors}, and \ref{fig:DME_scalar_reac_rate_recon_errors}. Notably, the exclusion of $\mathrm{N_2}$ as an inert species was not considered in this analysis. The results demonstrate that the overall nonlinear reconstruction employing ANN (blue bars) exhibits lower error compared to linear reconstruction (brown bars) across most species, temperature, production rates, and heat release rates for both PCA and CoK-PCA methods (Figures \ref{fig:DME_scalar_recon_errors}, \ref{fig:DME_reac_rate_recon_errors}).

\begin{figure}[h!]
    \centering
    {
    \includegraphics[width = 7.5cm]{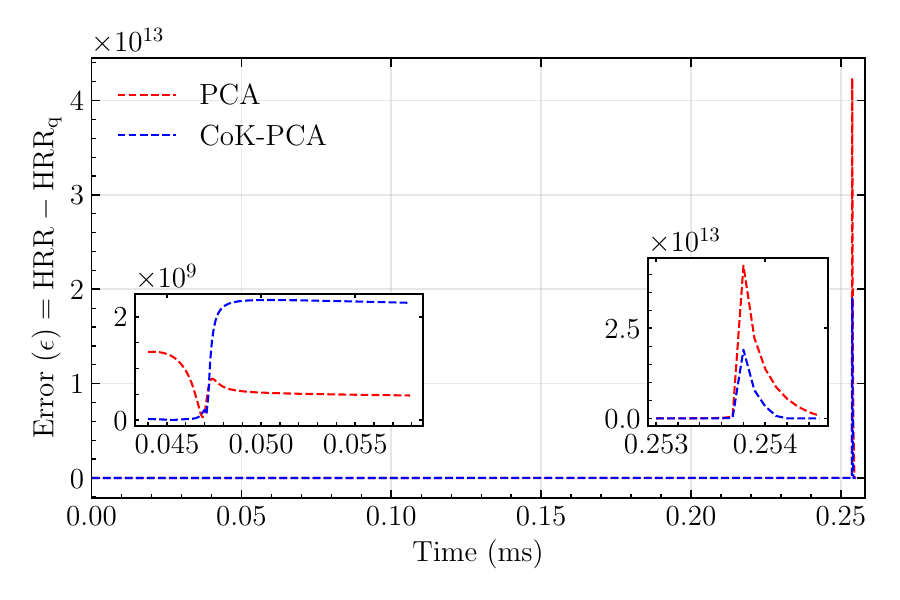}
    \begin{picture}(0,0)
        \put(-35,115){\scriptsize (a)}
    \end{picture}
    }
    {
    \includegraphics[width = 7.5cm]{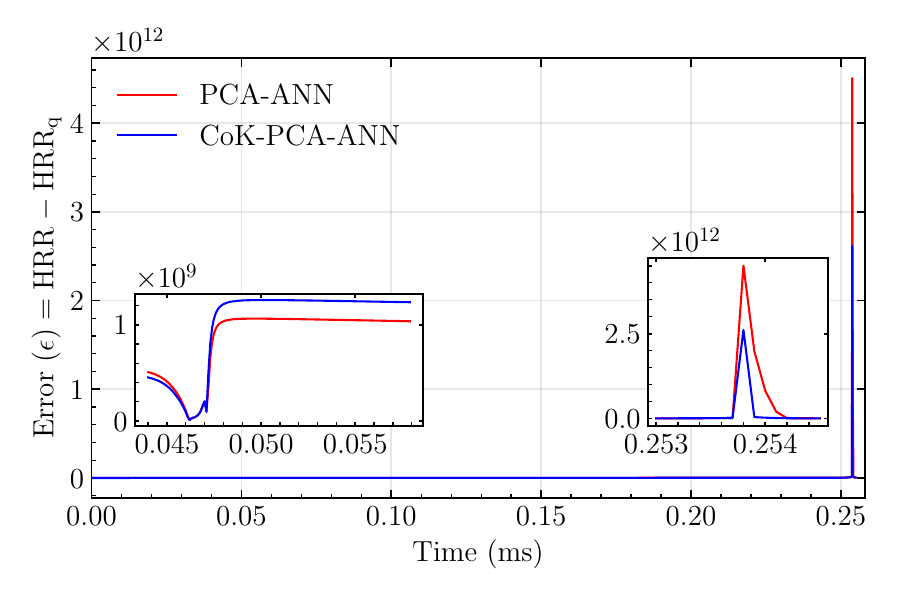}
    \begin{picture}(0,0)
        \put(-35,115){\scriptsize (b)}
    \end{picture}
    }
    \caption{Temporal evolution of absolute errors in the reconstruction of heat release rates for the DME two-stage autoignition flame with an initial temperature of 625 K.}
    \label{fig:DME_hrr_absolute_errors}
\end{figure}

Figure \ref{fig:DME_scalar_reac_rate_recon_errors} illustrates the error ratio between PCA-ANN and CoK-PCA-ANN for thermo-chemical scalars (\ref{fig:DME_scalar_reac_rate_recon_errors} (a) and (b)), species production rates, and heat release rates (\ref{fig:DME_scalar_reac_rate_recon_errors} (c) and (d)). The errors in reconstructed thermo-chemical scalars show mixed trends, unlike the ethylene-air dataset for which CoK-PCA-ANN was consistently more accurate than PCA-ANN. However, the accuracy of species production rates and, more importantly, the heat release rate for CoK-PCA-ANN is better than PCA-ANN. This result reinforces the notion that error metrics based only on thermo-chemical state reconstruction may not be sufficient measures of accuracy. Going beyond the error ratio, and similar to the ethylene-air case, we plot the absolute errors of heat release rate for one of the DME-air flames from the test set with an initial temperature of 625 K as shown in Fig.~\ref{fig:DME_hrr_absolute_errors}. Since this mixture has two-stage ignition, the heat release rate for the second stage (at $\sim$ 0.25 ms) is orders of magnitude larger than the first stage (at $\sim$ 0.047 ms). To make the comparison clearer, insets in Fig.~\ref{fig:DME_hrr_absolute_errors} show the regions zoomed on the two stages. It is evident that the absolute errors of heat release rate are greater by up to an order of magnitude with linear reconstruction (Fig.~\ref{fig:DME_hrr_absolute_errors} (a)) compared with ANN-based reconstruction (Fig.~\ref{fig:DME_hrr_absolute_errors} (b)). Moreover, while the errors for the first stage are comparable between PCA-ANN and CoK-PCA-ANN, for the second stage, CoK-PCA-ANN is more accurate.

\subsection{Premixed ethylene-air laminar flame}\label{sec:1D_ethylene_air}
The third case we consider is a one-dimensional freely-propagating planar laminar premixed flame of the ethylene-air mixture. In addition to the chemical reactions that govern the evolution of homogeneous reactors of the previous two cases, this case has effects of convection and diffusion that influence the thermo-chemical evolution. The chemistry is represented by the same 32-chemical species, 206-reactions mechanism~\cite{luo2012chemical}, resulting in $n_v = 33$ features. The freely-propagating flame is simulated in a one-dimensional domain of 0.02 m discretized with a grid of around 550 points. The pressure is kept at 1 atm, and a parametric variation is considered for the unburnt mixture conditions. Analogous to the ensemble training performed in Sec.~\ref{sec:0D_ethylene_air}, to construct the required training and testing data, we perturb the unburnt mixture temperature and equivalence ratio, (T, $\phi$), by $\Delta T = \pm 50 $ K and $\Delta \phi = \pm 0.25$ from the reference state, i.e., $D_1 \equiv (T = 300~\mathrm{K}, \phi = 0.6)$. This effectively results in nine configurations, $D_i~ \forall i \in \lbrace 1, 2, \cdots, 9 \rbrace$, one for each combination of (T, $\phi$) where T~$\in \lbrace 250~\mathrm{K},300~\mathrm{K},350~\mathrm{K} \rbrace$ and $\phi~\in \lbrace 0.575, 0.6, 0.625 \rbrace$. Again, to generate the CoK-PCA and PCA reduced manifolds, the principal components are computed with respect to the scaled reference state, $X_1$, by selecting $n_q = 5$ leading principal vectors out of the $n_v = 33$ vectors that capture approximately 99\% of the variance and 98\% of the kurtosis in the dataset, respectively. Following the dimensionality reduction procedure in Sec.~\ref{sec:dimensionality-reduction}, we compute the score matrices, $\mathbf{Z}^{4}_{q}$ and $\mathbf{Z}^{2}_{q}$ for the CoK-PCA and PCA low-dimensional manifolds, respectively.
\begin{figure}[h!]
    \centering
    {
    \includegraphics[width = 6cm]{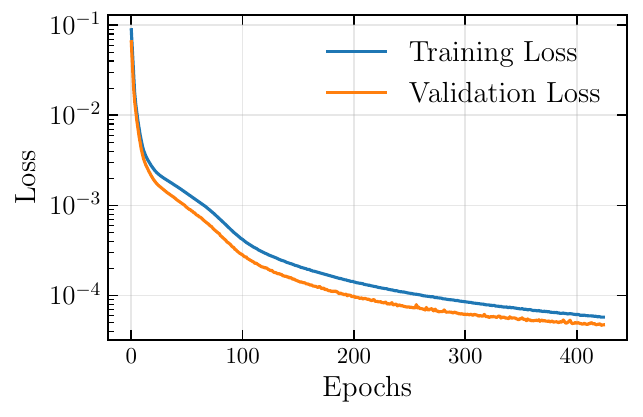}
    \begin{picture}(0,0)
        \put(-127,92){\scriptsize (a)}
    \end{picture}
    }
    {
    \includegraphics[width = 6cm]{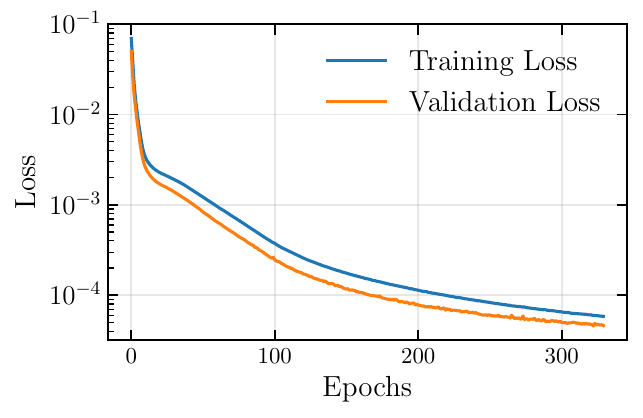}
    \begin{picture}(0,0)
        \put(-127,90){\scriptsize (b)}
    \end{picture}
    }
    \caption{Training and validation loss curves for (a) CoK-PCA-ANN and (b) PCA-ANN, respectively, for the one-dimensional planar laminar premixed ethylene-air flame dataset.}
    \label{fig:1D_9f_loss_curves}
\end{figure}

For the ANN training, a similar split of the data into training and testing sets, as Sec.~\ref{sec:0D_ethylene_air}, is performed here; $D_1$, $D_2$, $D_4$, $D_6$, and $D_8$ are used for training and the rest for testing. Accordingly, we construct the input feature vectors and ground truths to train a neural network with four hidden layers of widths 48, 48, 48, and 56 neurons. The widths of input and output layers are $n_q = 5$ and $n_v = 33$ neurons, respectively. The layer activation functions remain the same as before, a hyperbolic tangent function,  with the use of Adam optimizer (learning rate = \num{1e-4}) for training. Figures~\ref{fig:1D_9f_loss_curves} (a) and (b) depict the loss curves obtained for CoK-PCA-ANN and PCA-ANN, respectively.
\begin{figure*}[h!]
    \centering
    {
    \includegraphics[width = 7.98cm]{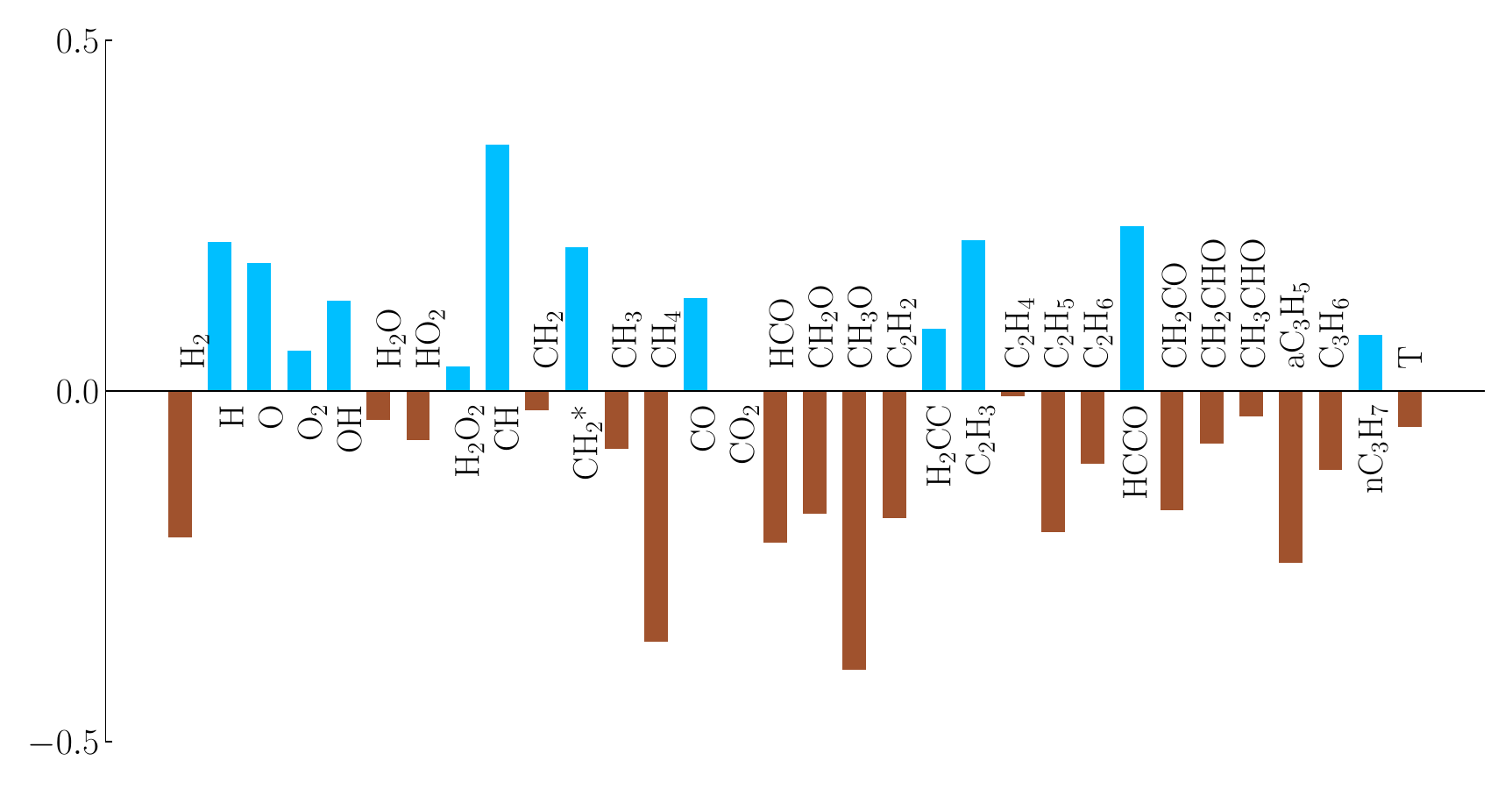}
    \begin{picture}(0,0)
        \put(-208,108){\scriptsize(a) Species mass fractions and temperature}
        \put(-237,50){\scriptsize {\rotatebox{90}{Error ratio $r_a$}}}
    \end{picture}
    }\hspace{-0.018cm}
    {
    \includegraphics[width = 7.98cm]{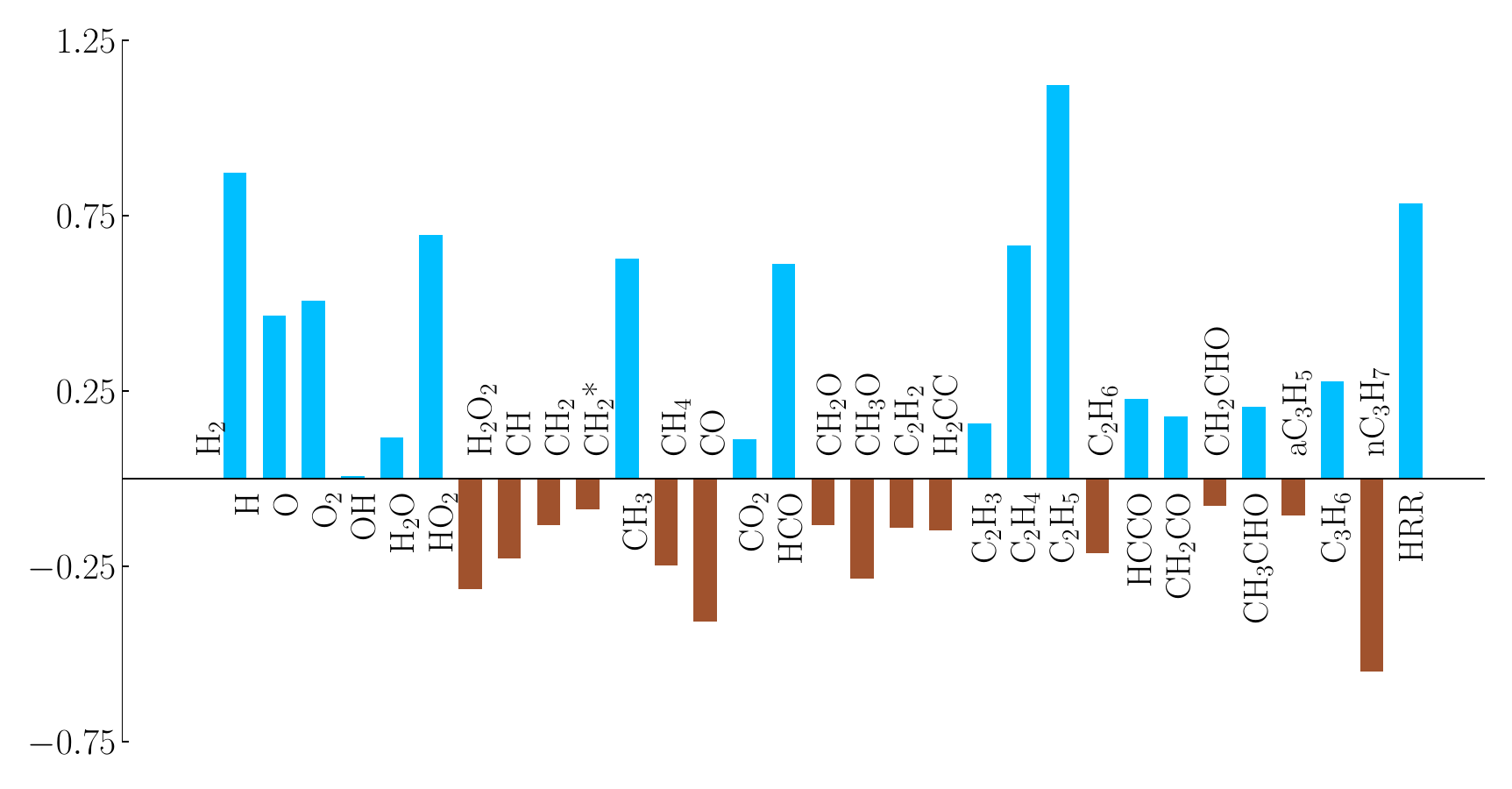}
    \begin{picture}(0,0)
        \put(-208,108){\scriptsize(c) Species production rates and heat release rate}
        \put(-237,50){\scriptsize {\rotatebox{90}{Error ratio $r_a$}}}
    \end{picture}
    }
    {
    \includegraphics[width = 7.98cm]{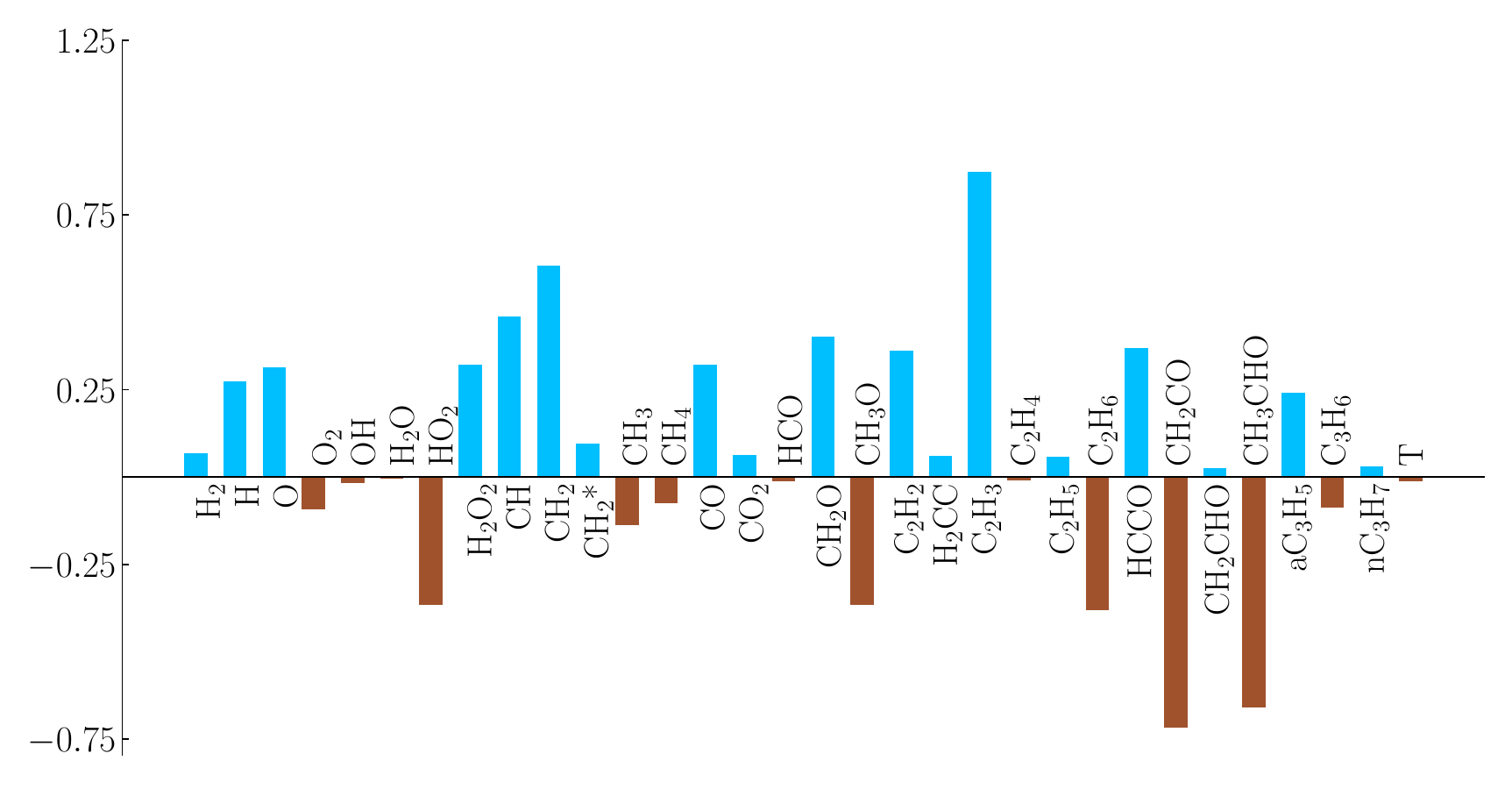}
    \begin{picture}(0,0)
        \put(-206,113){\scriptsize(b) Species mass fractions and temperature}
        \put(-236,50){\scriptsize {\rotatebox{90}{Error ratio $r_m$}}}
    \end{picture}
    }\hspace{-0.018cm}
    {
    \includegraphics[width = 7.98cm]{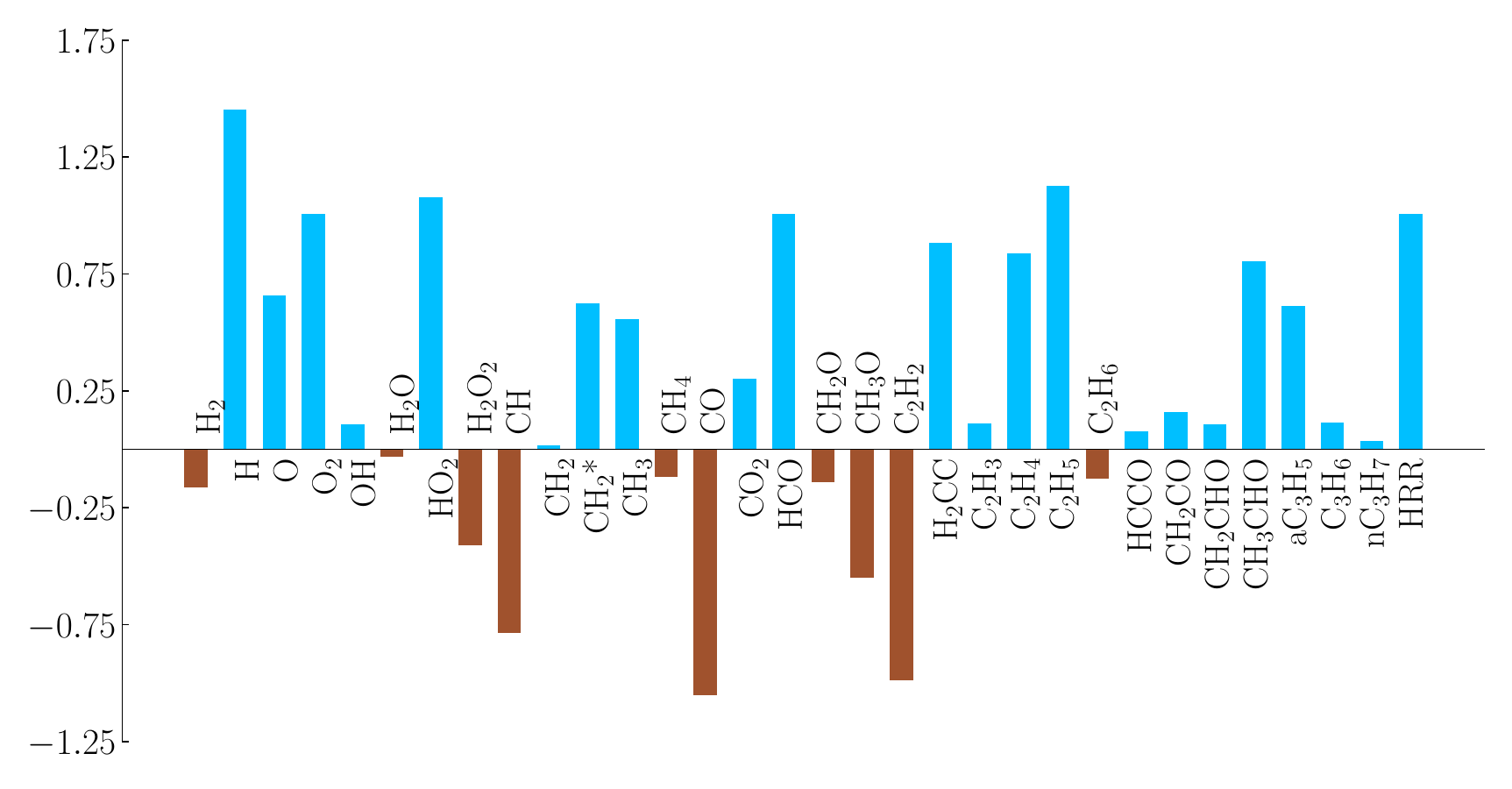}
    \begin{picture}(0,0)
        \put(-206,113){\scriptsize(d) Species production rates and heat release rate}
        \put(-236,50){\scriptsize {\rotatebox{90}{Error ratio $r_m$}}}
    \end{picture}
    }
    \caption{Comparison of errors in the reconstruction of thermo-chemical scalars (left), species production rates and heat release rate (right) with PCA-ANN vs. CoK-PCA-ANN for the one-dimensional planar laminar premixed ethylene-air flame dataset. Top and bottom plots in each column represent $r_a$ and $r_m$ respectively.}
    \label{fig:1D_9f_nonlinear_pca_cok-pca}
\end{figure*}

Following Sec.~\ref{sec:0D_ethylene_air}, we assess the reconstruction accuracy of the trained models on the test states, i.e., $D_j~\forall j \in \lbrace 3, 5, 7, 9 \rbrace$. Similar to the trends observed in previous cases, ANN reconstruction outperforms linear reconstruction for all the quantities of interest, the plots of which are not presented here for brevity. With reconstruction based on ANNs, we next focus on the performance of CoK-PCA-ANN against PCA-ANN in terms of the error ratios ($r_a$, $r_m$), which are presented in Fig.~\ref{fig:1D_9f_nonlinear_pca_cok-pca}. For the accuracy of thermo-chemical scalars, we observe a different trend in this case, with PCA-ANN being more accurate than CoK-PCA-ANN for 19 out of the 33 variables for $r_a$. However, as hypothesized, CoK-PCA-ANN performs better than PCA-ANN in terms of the $r_m$ metric in accurate predictions of 21 out of the 33 variables. Further, while comparing errors in the reconstruction of species production rates and heat release rates, CoK-PCA-ANN dominates over PCA-ANN in both error ratios. In particular, CoK-PCA-ANN significantly improves upon PCA-ANN by predicting production rates for 22 out of 32 species in terms of the $r_m$ error and 18 out of 32 species in terms of the $r_a$ error. More importantly, it incurs lower errors in reconstructing the heat release rate in both metrics, which is an overall measure of the fidelity of the chemical system. This case clearly illustrates the fact that errors in reconstructing the thermo-chemical state alone might not be a sufficient measure of accuracy for a given dimensionality reduction technique, and a broader set of metrics might be prudent.
\begin{figure*}[h!]
    \centering
    {
    \includegraphics[width=6.5cm]{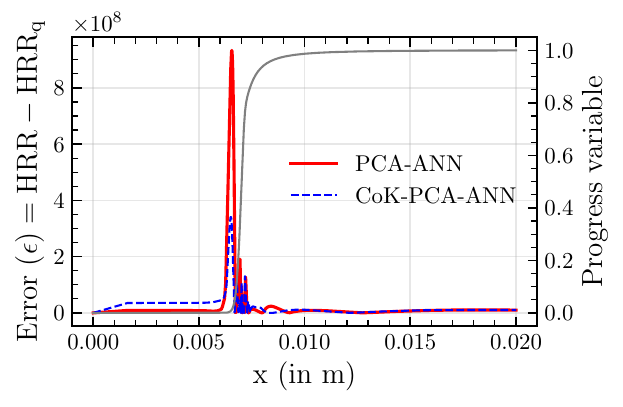}
    \begin{picture}(0,0)
        \put(-153,95){\scriptsize(a)}
    \end{picture}
    }
    \hspace{0.2cm}
    {
    \includegraphics[width=6.5cm]{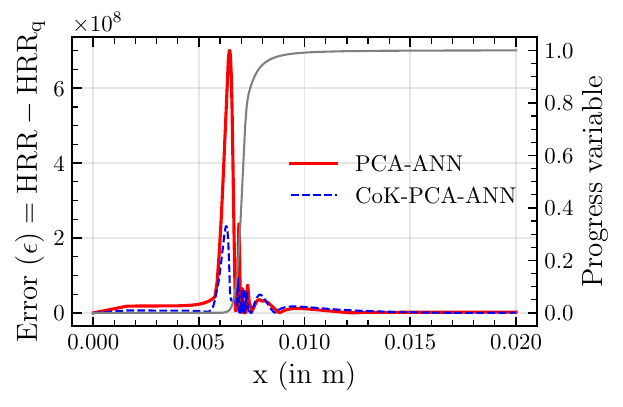}
    \begin{picture}(0,0)
        \put(-153,95){\scriptsize(c)}
    \end{picture}
    }
    \vspace{-0.3cm}
    {
    \includegraphics[width=6.5cm]{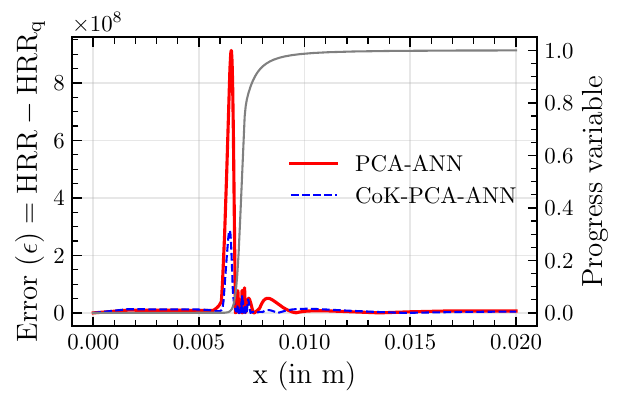}
    \begin{picture}(0,0)
        \put(-152,93){\scriptsize(b)}
    \end{picture}
    }
    \hspace{0.2cm}
    {
    \includegraphics[width=6.5cm]{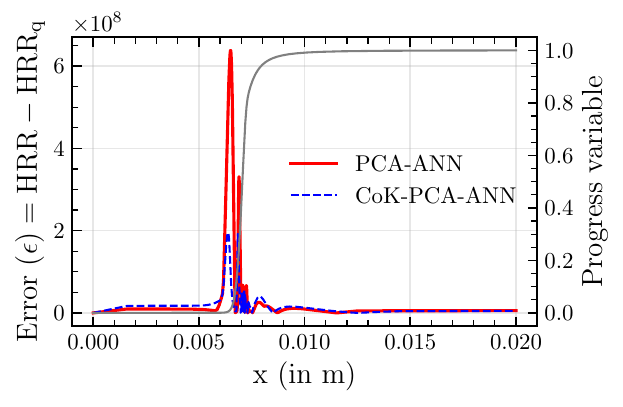}
    \begin{picture}(0,0)
        \put(-157,95){\scriptsize(d)}
    \end{picture}
    }
    \caption{Spatial variation of absolute errors in reconstructed heat release rates for the test states - (a) $D_3$, (b) $D_5$, (c) $D_7$, and (d) $D_9$ for the one-dimensional planar laminar premixed ethylene-air flame dataset. The progress variable is plotted in grey for reference.}
    \label{fig:1D_9f_cok-ann-pca-ann_hrr}
\end{figure*}

The profile of absolute errors in heat release rates obtained for both the methods, CoK-PCA-ANN (dashed blue) and PCA-ANN (solid red), is shown in Fig.~\ref{fig:1D_9f_cok-ann-pca-ann_hrr} for the four test states, $D_3$, $D_5$, $D_7$, and $D_9$. We observe that CoK-PCA-ANN outperforms PCA-ANN in accurately predicting the steady-state flame location for all the test states, thereby characterizing flame propagation better. This behavior is consistent with the $r_m$ errors presented in Fig.~\ref{fig:1D_9f_nonlinear_pca_cok-pca} (d). Further, both techniques capture the non-reacting regions reasonably well in all the test states. However, in these regions, CoK-PCA-ANN performs marginally better than PCA-ANN by predicting nearly zero heat release rates for the test flames, $D_5, D_7,$ and $D_9$ (Figures~\ref{fig:1D_9f_cok-ann-pca-ann_hrr} (b), (c), (d)). It should be noted that negligible reconstruction errors incurred by the methods in these regions (i.e., predicting non-zero heat release in the non-reacting zones) can be attributed to statistical inconsistencies or stochasticity of the ANN training process. Consequently, this is reflected in the $r_a$ metric (average error), which is lesser in the case of CoK-PCA-ANN than PCA-ANN (demonstrated by blue bars) in Fig.~\ref{fig:1D_9f_nonlinear_pca_cok-pca} (c). 

\subsection{Homogeneous charge compression ignition}
\label{sec:2D_hcci}
\begin{figure}
    \centering
    \includegraphics[width = 6cm]{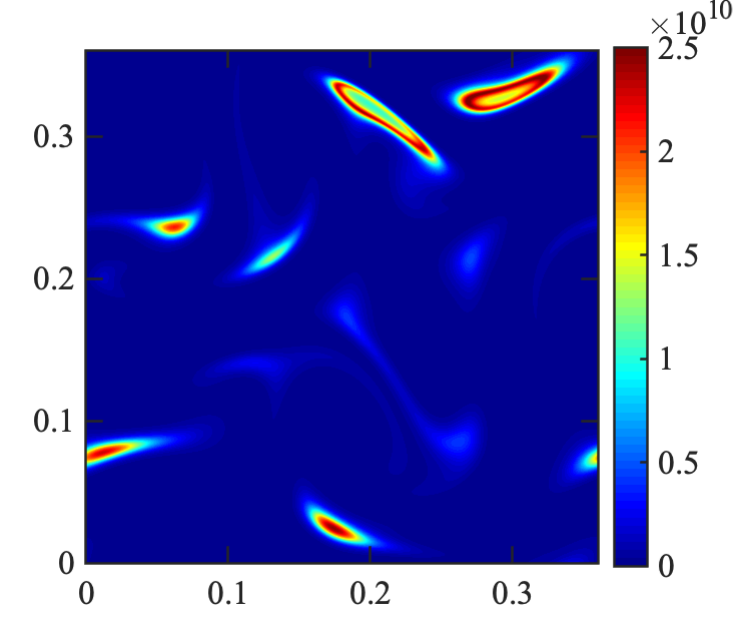}
    \caption{Instantaneous contour plots of heat release rates ($\mathrm{Jm^{-3}s^{-1}}$) from two-dimensional HCCI dataset at time (t) = 1.2 ms.}
    \label{fig:hrr_original_1.2ms}
\end{figure}

In this section, we examine a dataset that encompasses the influence of spatial transport involving convection and diffusion and turbulence. The dataset focuses on homogeneous charge compression ignition (HCCI) of ethanol, which is representative of internal combustion engines~\cite{bhagatwala-hcci-dns}. The simulation corresponds to high-pressure, high-temperature auto-ignition of a turbulent mixture composed of premixed ethanol-air and combustion products, emulating the process of ``exhaust gas recirculation'' (EGR). The simulation is performed in a fully periodic domain with a two-dimensional spatial grid of 672 × 672 points. The initial conditions include a nominal pressure of \SI{45}{atm} and a mean temperature of \SI{924}{K}. The reactants are set to an equivalence ratio of 0.4. To account for the uneven mixing caused by EGR, a spatial temperature fluctuation and a separately computed divergence-free turbulent velocity field are superimposed onto the system. Furthermore, the simulation also considers the effects of compression heating resulting from the motion of the piston. The chemistry is represented by a 28 chemical species reaction mechanism. Thus, at each simulation snapshot, the design matrix, $\mathbf{D}$ consists of $n_g = 672 \times 672$ data samples and $n_v = 29$ thermo-chemical scalars. For this study, we consider the temporal checkpoint at t = 1.2 ms~\cite{aditya-anomaly-detection-2019-JCP}, which corresponds to the propagation of the flame fronts in the bulk of the global domain, as shown in the heat release rate contours in Fig.~\ref{fig:hrr_original_1.2ms}, which has been saturated to a peak heat release rate of $1 \times 10^{9}~\mathrm{Jm^{-3}s^{-1}}$ in order to demonstrate the growth in the size of the ignition kernels.
\begin{figure}[h!]
    \centering
    {
    \includegraphics[width = 6cm]{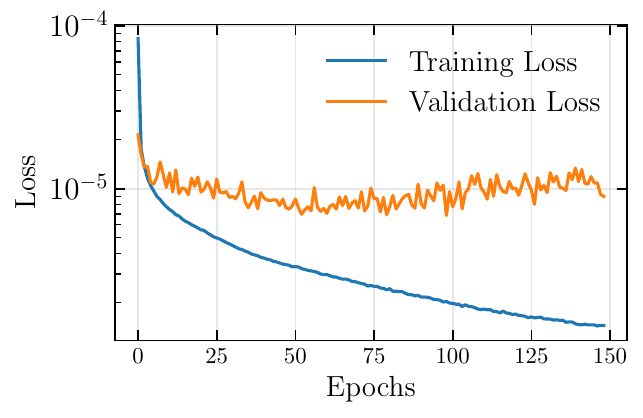}
    \begin{picture}(0,0)
        \put(-130,90){\scriptsize (a)}
    \end{picture}
    }
    {
    \includegraphics[width = 6cm]{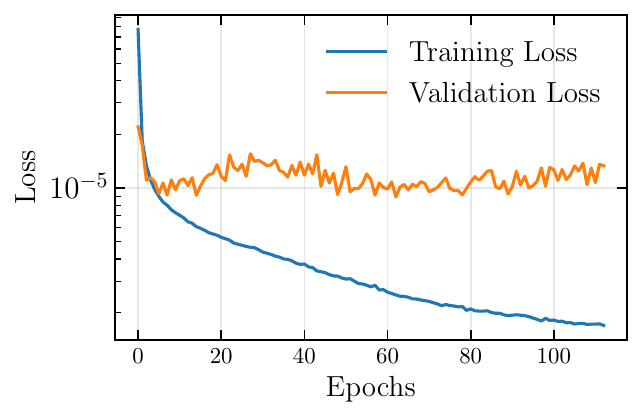}
    \begin{picture}(0,0)
        \put(-130,90){\scriptsize (b)}
    \end{picture}
    }
    \caption{Training and validation loss curves for - (a) CoK-PCA-ANN and (b) PCA-ANN, respectively, for the two-dimensional HCCI dataset.}
    \label{fig:2D_hcci_loss_curves}
\end{figure}

For the testing state, we consider the simulation snapshot at 1.19 ms. In other words, we are interested in investigating the efficacy of the proposed CoK-PCA-ANN method in predicting the thermo-chemical state at an unseen state (t = 1.19 ms) while being trained on a subsequent checkpoint at t = 1.2 ms. To obtain the score matrices $\mathbf{Z}^4_q$ and $\mathbf{Z}^2_q$, we use the principal vectors computed on the reference state, i.e., on t = 1.2 ms. The low-dimensional manifolds are constructed by retaining $n_q = 5$ out of $n_v = 29$ principal vectors that correspond to approximately 99\% of the variance and kurtosis in the reduced PCA and CoK-PCA manifolds, respectively. The next step involves constructing the required train and test data comprising input feature vectors and corresponding ground truths. It should be noted that since this is a two-dimensional dataset, we suitably flatten it to yield $N = 451584$ data samples. Further, a neural network with three hidden layers of widths 8, 8, and 64 neurons is trained till convergence with an Adam optimizer (learning rate = 0.028). In addition, early stopping is employed to ensure the network does not lead to overfitting of the training data. The corresponding loss curves obtained for the manifolds are presented in Fig.~\ref{fig:2D_hcci_loss_curves}. We then use the trained network to predict the thermo-chemical scalars at t = 1.19 ms for both CoK-PCA and PCA reduced manifolds. In a similar manner, using the reconstructed thermo-chemical scalars, species production rates and heat release rates are computed. In the following, we present a comparison of CoK-PCA-ANN and PCA-ANN in terms of the reconstruction errors of the aforementioned quantities.
\begin{figure*}[h!]
    \centering
    {
    \includegraphics[width = 7.98cm]{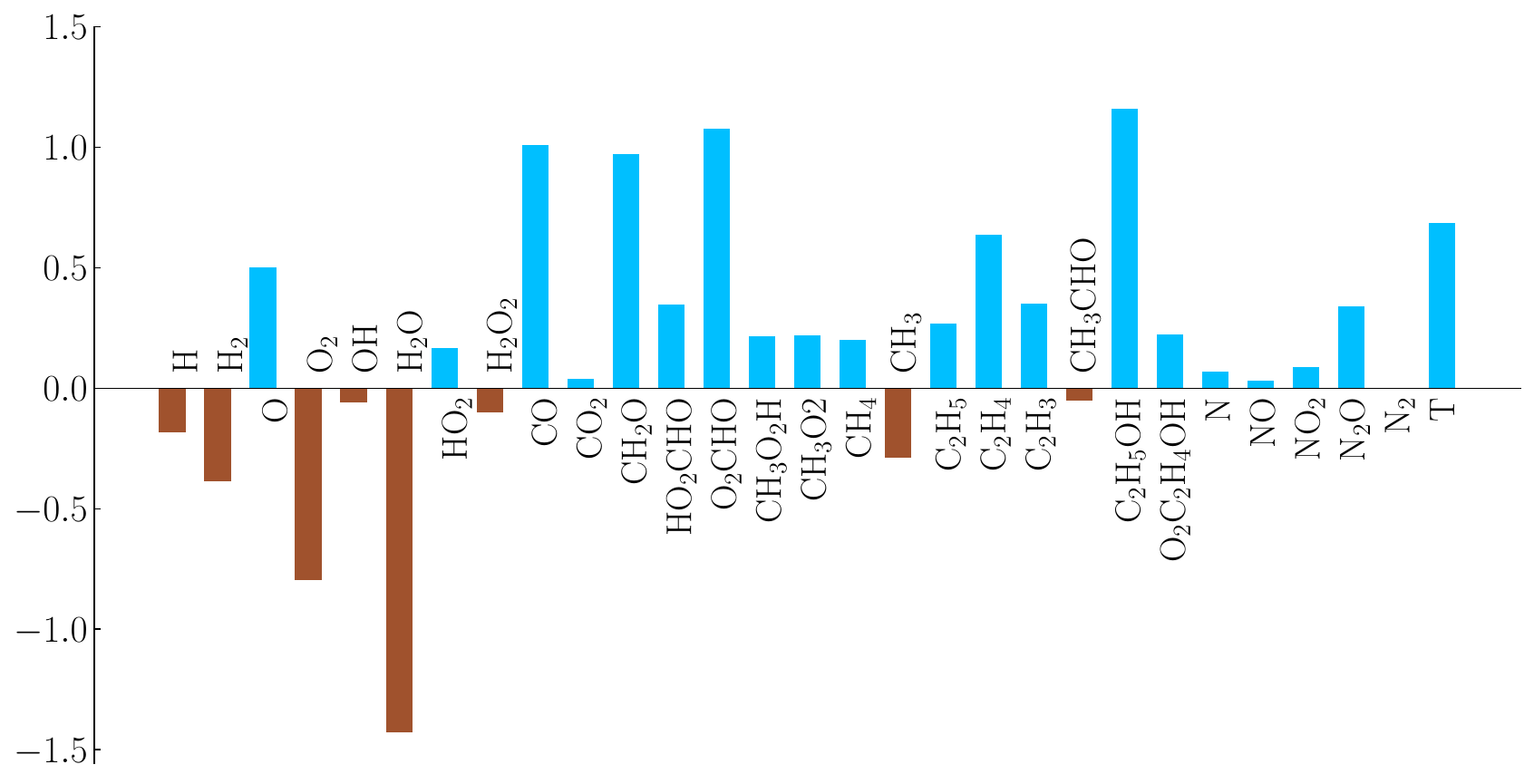}
    \begin{picture}(0,0)
        \put(-208,108){\scriptsize(a) Species mass fractions and temperature}
        \put(-237,50){\scriptsize {\rotatebox{90}{Error ratio $r_a$}}}
    \end{picture}
    }\hspace{-0.018cm}
    {
    \includegraphics[width=7.98cm]{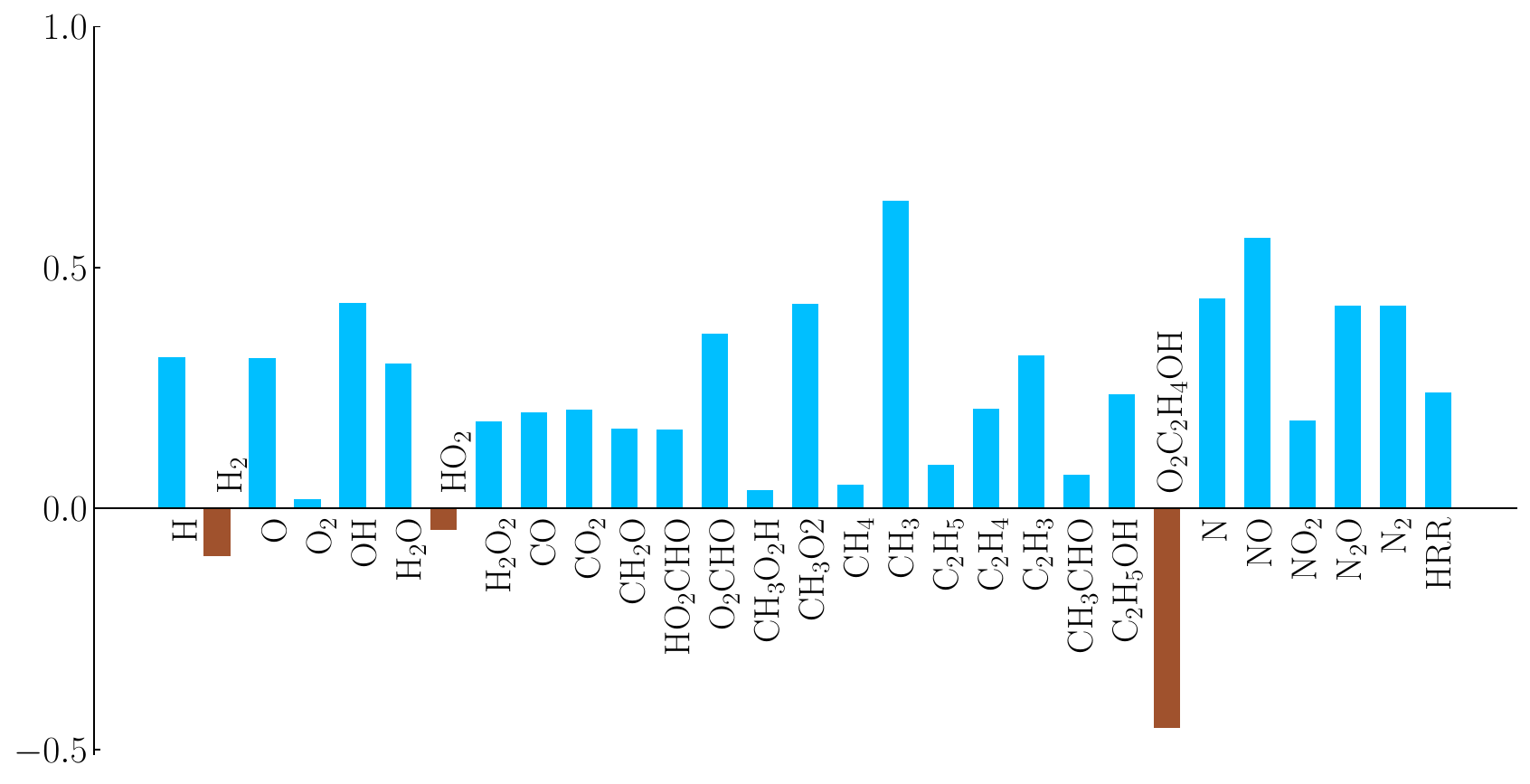}
    \begin{picture}(0,0)
        \put(-208,108){\scriptsize(c) Species production rates and heat release rate}
        \put(-237,50){\scriptsize {\rotatebox{90}{Error ratio $r_a$}}}
    \end{picture}
    }
    {
    \includegraphics[width=7.98cm]{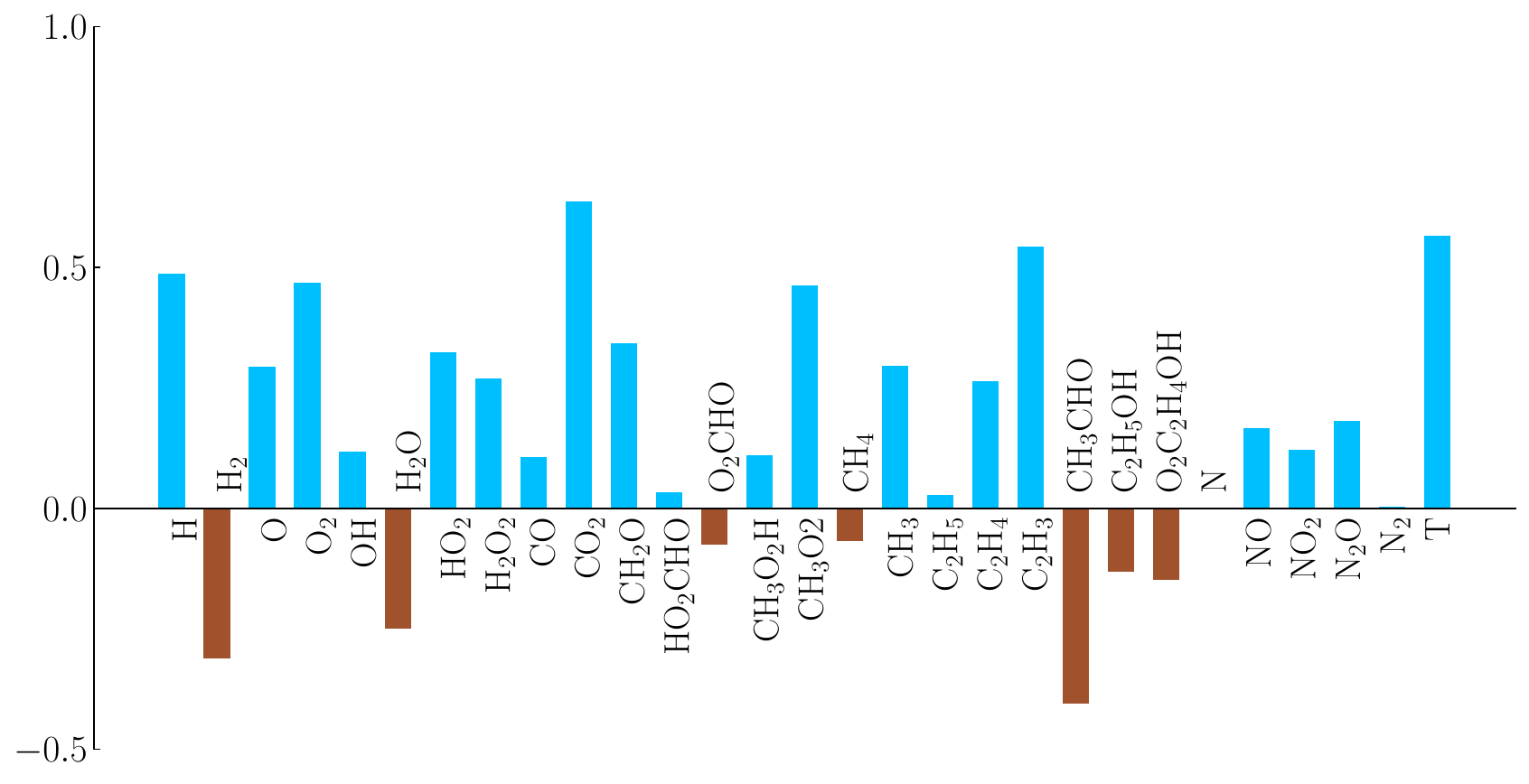}
    \begin{picture}(0,0)
        \put(-206,108){\scriptsize(b) Species mass fractions and temperature}
        \put(-236,50){\scriptsize {\rotatebox{90}{Error ratio $r_m$}}}
    \end{picture}
    }\hspace{-0.018cm}
    {
    \includegraphics[width=7.98cm]{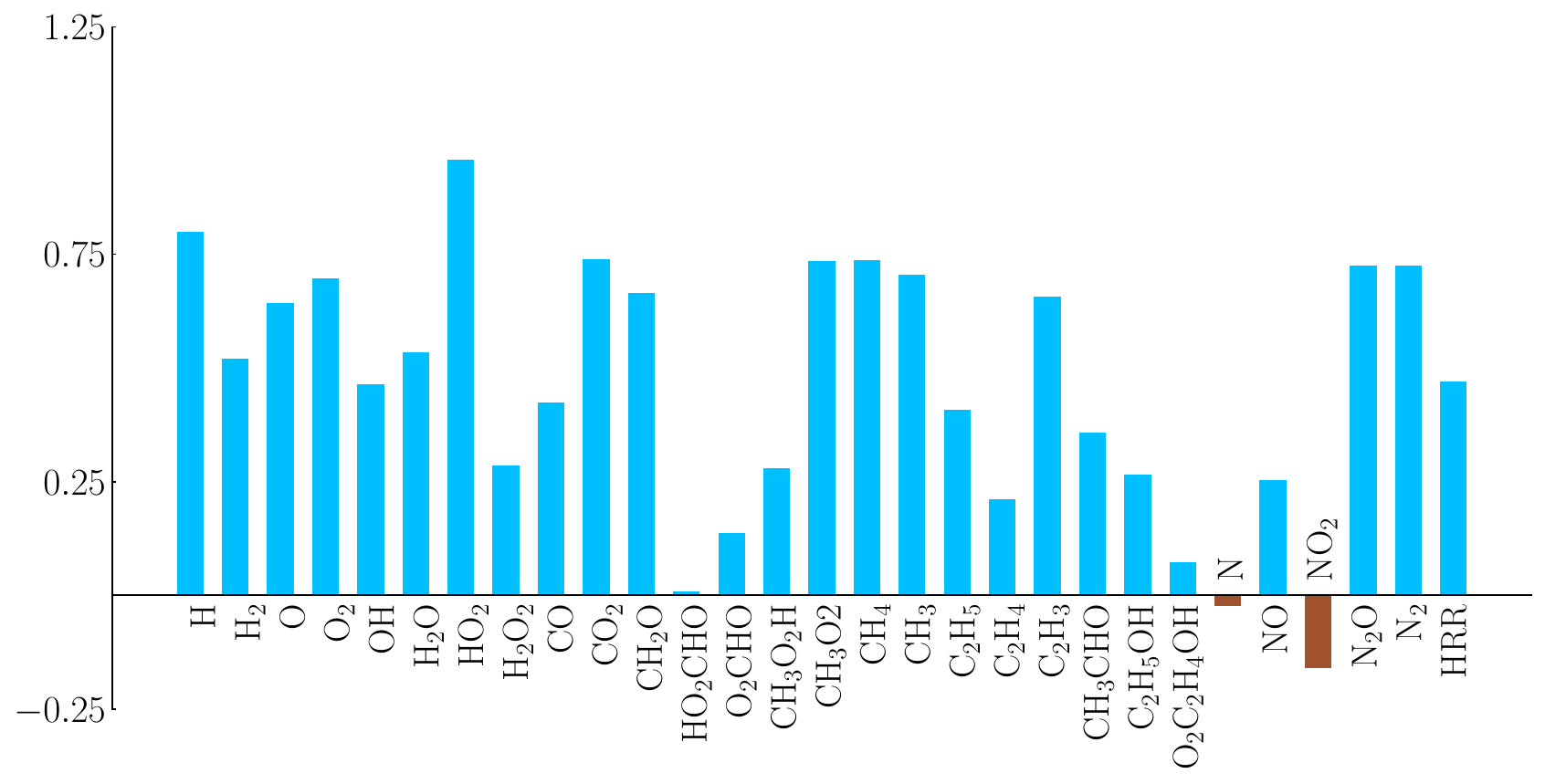}
   \begin{picture}(0,0)
        \put(-206,108){\scriptsize(d) Species production rates and heat release rate}
        \put(-236,50){\scriptsize {\rotatebox{90}{Error ratio $r_m$}}}
    \end{picture}
    }
    \caption{Comparison of errors in the reconstruction of thermo-chemical scalars (left), species production rates and heat release rate (right) for PCA-ANN vs. CoK-PCA-ANN for the two-dimensional HCCI dataset. Top and bottom plots in each column represent $r_a$ and $r_m$ respectively.}
    \label{fig:2D_hcci_nonlinear_pca_cok-pca}
\end{figure*}

\begin{figure*}[h!]
    \centering
    {
    \includegraphics[width=5.37cm]{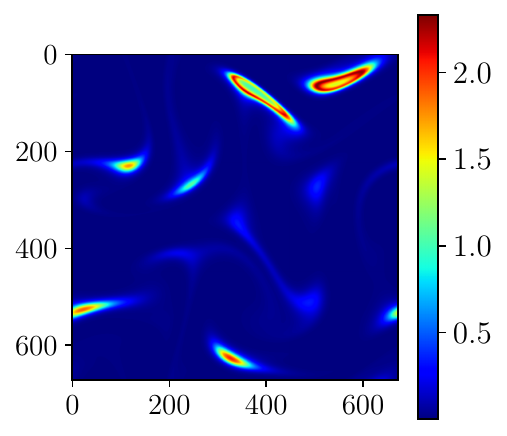}
    \begin{picture}(0,0)
        \put(-88,-10){\scriptsize (a)}
    \end{picture}
    }\hspace{-0.2cm}
    {
    \includegraphics[width=5.37cm]{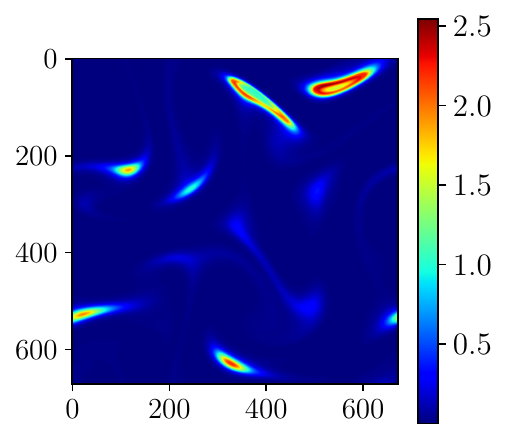}
    \begin{picture}(0,0)
        \put(-88,-10){\scriptsize (b)}
    \end{picture}
    }\hspace{-0.2cm}
    {
    \includegraphics[width=5.37cm]{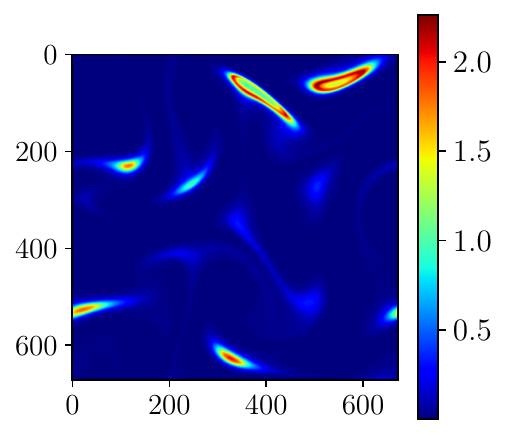}
    \begin{picture}(0,0)
        \put(150,2){\scriptsize (c)}
    \end{picture}
    }
    \caption{Instantaneous contour plots of heat release rates ($\mathrm{Jm^{-3}s^{-1}}$) for the two-dimensional HCCI dataset at t = 1.19 ms. The columns (left to right) represent the heat release rates from the (a) original, (b) PCA reconstructed, and (c) CoK-PCA reconstructed datasets, respectively.}
    \label{fig:2D_hcci_hrr_contours} 
\end{figure*}

In Fig.~\ref{fig:2D_hcci_nonlinear_pca_cok-pca}, it is evident that CoK-PCA-ANN performs significantly better than PCA-ANN in the reconstruction of thermo-chemical scalars with more accurate predictions of 20 out of 29 species in both $r_a$ and $r_m$ errors. Furthermore, it completely dominates PCA-ANN in accurately reconstructing production rates for 90\% and 93\% of the species in terms of $r_a$ and $r_m$ metrics, respectively. In addition, it provides an accurate representation of the chemical dynamics in the reaction zones by incurring lower reconstruction errors for heat release rates in both metrics ($r_m, r_a$). Contrary to the observations in~\cite{jonnalagadda2023co}, where the CoK-PCA-based manifold performed poorly in terms of the average errors ($r_a$) while considering the entire spatial domain, CoK-PCA, when coupled with ANN, overcomes this issue and better represents the stiff chemical dynamics in the average error as well. Next, we plot the reconstructed heat release rate contours in Fig.~\ref{fig:2D_hcci_hrr_contours} to get a qualitative difference in the magnitude of the heat release rate within the ignition kernels achievable through CoK-PCA and PCA-reduced manifolds. Due to the inherent ability of excess kurtosis to suitably capture outliers, CoK-PCA-ANN identifies the ignition zones better (as demonstrated by the better matching of color hues with the original) in the entire domain, which is in good agreement with the lesser maximum error ($r_m$) as shown in Fig.~\ref{fig:2D_hcci_nonlinear_pca_cok-pca} (d) for heat release rate. Additionally, due to the coupling of ANN with CoK-PCA, the non-igniting regions are also represented better than PCA-ANN, leading to a lower average error ($r_a$) as presented in Fig.~\ref{fig:2D_hcci_nonlinear_pca_cok-pca} (c).     

\section{Conclusions and future work}\label{sec:conclusions} 
In this paper, we have proposed an enhanced version of the co-kurtosis PCA (CoK-PCA) based dimensionality reduction method, namely CoK-PCA-ANN, which leverages the potential of artificial neural networks (ANNs) to model complex nonlinear relationships inherent between the aggressively truncated low-dimensional manifolds and the original thermo-chemical state. The rationale behind this work is to assess the overall efficacy of the CoK-PCA method in comparison to PCA in conjunction with nonlinear reconstruction methods and expand its applicability to chemically reacting systems presenting stiff dynamics. A brief overview of the various state-of-the-art nonlinear reconstruction methods, such as ANNs, gaussian process regression (GPR), kernel density methods, autoencoders, etc., combined with PCA was discussed. In contrast, these methods are yet to be explored in the CoK-PCA framework which motivates this work. The framework of the proposed CoK-PCA-ANN dimensionality reduction method was presented with a discussion on the generation of the low-dimensional manifold using linear projection (encoding) with CoK-PCA followed by nonlinear reconstruction of the original thermochemical state space (decoding) using ANNs. The performance of the CoK-PCA-ANN method was benchmarked with linear CoK-PCA and PCA-ANN methods for four combustion test cases that characterize various physical and chemical phenomena in reacting flows (e.g., autoignition, flame propagation): a homogeneous reactor simulation representing conventional single-stage and complex two-stage autoignition, a one-dimensional freely propagating laminar premixed flame exhibiting flame propagation, and two-dimensional turbulent autoignition in homogeneous charge compression ignition conditions. In contrast to linear methods, ANNs demonstrated significantly high reconstruction accuracies for the CoK-PCA and PCA manifolds in terms of thermo-chemical scalars, species production rates, and heat release rates with aggressive truncation (low $n_q$). Further, the quality of the manifolds was assessed in conjunction with ANN for the aforementioned quantities of interest. As hypothesized, CoK-PCA-ANN outperforms PCA-ANN in all the test cases in terms of the maximum $r_m$ errors for the thermo-chemical scalars, species production rates, and most importantly, heat release rates, thereby reinforcing the fact that the chemical kinetics prevalent in the ignition zones representative of stiff dynamics is captured more accurately by CoK-PCA than PCA. Contrary to the findings in the previous assessment of plain vanilla CoK-PCA~\cite{jonnalagadda2023co}, CoK-PCA-ANN incurred lower reconstruction errors in the average error metric ($r_a$) as well with a better representation of the unburnt reactants and burnt products in all the test cases. Additionally, CoK-PCA-ANN outperforms PCA-ANN in accurately predicting an unseen test state different from the training set considered in the 2D HCCI case. To summarize, the results from the above analyses suggest that CoK-PCA-ANN realizes the advantages of both CoK-PCA and ANNs and proves reliable, robust, and generalizable to unseen thermo-chemical states that share similar ignition kinetics as the training state. 

However, it should be remarked that, in this paper, the investigation of CoK-PCA-based nonlinear reconstruction using ANNs was carried out in an \textit{a priori} setting. It is well known that these data-driven dimensionality reduction methods are capable of accelerating numerical simulations of reacting flows by solving a reduced set of principal component transport equations as opposed to solving a very high-dimensional system of species conservation equations. Such a kind of \textit{a posteriori} validation performed for PCA remains to be explored for CoK-PCA and therefore forms the future scope of this paper.

\section*{Acknowledgments}
The work at IISc was supported under a project from the National Supercomputing Mission, India. DN is a recipient of the Ansys M.Tech. (Research) Fellowship. AJ was funded by a project from Shell Technology Center, Bengaluru, India. KA is a recipient of the Arcot Ramachandran Young Investigator award, IISc.
Work by HK and UB was part of the ExaLearn Co-design Center, supported by the
Exascale Computing Project (17-SC-20-SC), a collaborative effort of the U.S. Department of Energy Office of Science and the National Nuclear Security Administration.
Sandia National Laboratories is a multi-mission laboratory managed and operated by National Technology and Engineering Solutions of Sandia, LLC., a wholly owned subsidiary of Honeywell International, Inc., for the U.S. Department of Energy’s National Nuclear Security Administration under contract DE-NA-0003525.
The views expressed in the article do not necessarily represent the views of the U.S. Department of Energy or the United States Government.

\bibliographystyle{elsarticle-num}
\bibliography{dimred.bib}

\begin{thebibliography}{10}
\expandafter\ifx\csname url\endcsname\relax
  \def\url#1{\texttt{#1}}\fi
\expandafter\ifx\csname urlprefix\endcsname\relax\def\urlprefix{URL }\fi
\expandafter\ifx\csname href\endcsname\relax
  \def\href#1#2{#2} \def\path#1{#1}\fi

\bibitem{adityaDirectNumericalSimulation2019}
K.~Aditya, A.~Gruber, C.~Xu, T.~Lu, A.~Krisman, M.~R. Bothien, J.~H. Chen,
  Direct numerical simulation of flame stabilization assisted by autoignition
  in a reheat gas turbine combustor, Proc. Combust. Inst. 37 (2019) 2635--2642.

\bibitem{savard2019}
B.~Savard, E.~R. Hawkes, K.~Aditya, H.~Wang, J.~H. Chen, Regimes of premixed
  turbulent spontaneous ignition and deflagration under gas-turbine reheat
  combustion conditions, Combust. Flame 208 (2019) 402--419.

\bibitem{bergerDNSStudyImpact2020}
L.~Berger, R.~Hesse, K.~Kleinheinz, M.~J. Hegetschweiler, A.~Attili,
  J.~Beeckmann, G.~T. Linteris, H.~Pitsch, A {{DNS}} study of the impact of
  gravity on spherically expanding laminar premixed flames, Combust. Flame 216
  (2020) 412--425.

\bibitem{Nivarti2017}
G.~Nivarti, S.~Cant, Direct numerical simulation of the bending effect in
  turbulent premixed flames, Proc. Combust. Inst. 36 (2017) 1903--1910.

\bibitem{desai2021direct}
S.~Desai, Y.~J. Kim, W.~Song, M.~B. Luong, F.~E.~H. P{\'e}rez, R.~Sankaran,
  H.~G. Im, Direct numerical simulations of turbulent reacting flows with shock
  waves and stiff chemistry using many-core/gpu acceleration, Comput. Fluids
  215 (2021) 104787.

\bibitem{uranakara2022accelerating}
H.~A. Uranakara, S.~Barwey, F.~E.~H. P{\'e}rez, V.~Vijayarangan, V.~Raman,
  H.~G. Im, Accelerating turbulent reacting flow simulations on many-core/gpus
  using matrix-based kinetics, Proceedings of the Combustion Institute (2022).

\bibitem{sutherland-parante-2009}
J.~C. Sutherland, A.~Parente, Combustion modeling using principal component
  analysis, Proc. Combust. Inst. 32 (2009) 1563--1570.

\bibitem{biglari-sutherland-2012}
A.~Biglari, J.~C. Sutherland, A filter-independent model identification
  technique for turbulent combustion modeling, Combust. Flame 159 (2012)
  1960--1970.

\bibitem{yang-pope-chen-2013}
Y.~Yang, S.~B. Pope, J.~H. Chen, Empirical low-dimensional manifolds in
  composition space, Combust. Flame 160 (2013) 1967--1980.

\bibitem{ranade-echeckki-2019}
R.~Ranade, T.~Echekki, A framework for data-based turbulent combustion closure:
  A priori validation, Combust. Flame 206 (2019) 490--505.

\bibitem{parente-sutherland-dally-tognotti-smith-localPCA-MILD-2011}
A.~Parente, J.~Sutherland, B.~B. Dally, L.~Tognotti, P.~Smith, Investigation of
  the mild combustion regime via principal component analysis, Proc. Combust.
  Inst. 33 (2011) 3333--3341.

\bibitem{parente2009identification}
A.~Parente, J.~C. Sutherland, L.~Tognotti, P.~J. Smith, Identification of
  low-dimensional manifolds in turbulent flames, Proceedings of the Combustion
  Institute 32~(1) (2009) 1579--1586.

\bibitem{aditya-anomaly-detection-2019-JCP}
K.~Aditya, H.~Kolla, W.~P. Kegelmeyer, T.~M. Shead, J.~Ling, W.~L. Davis,
  Anomaly detection in scientific data using joint statistical moments, J.
  Comput. Phys. 387 (2019) 522--538.

\bibitem{jonnalagadda2023co}
A.~Jonnalagadda, S.~Kulkarni, A.~Rodhiya, H.~Kolla, K.~Aditya, A co-kurtosis
  based dimensionality reduction method for combustion datasets, Combustion and
  Flame 250 (2023) 112635.

\bibitem{chen2023direct}
J.~Chen, S.~Desai, H.~Babaee, S.~Yamajala, Direct numerical simulation with
  time dependent subspaces for reduced-order modeling (rom) of turbulent
  compressible reacting flows, Bulletin of the American Physical Society
  (2023).

\bibitem{mirgolbabaei2015reconstruction}
H.~Mirgolbabaei, T.~Echekki, The reconstruction of thermo-chemical scalars in
  combustion from a reduced set of their principal components, Combust. Flame
  162~(5) (2015) 1650--1652.

\bibitem{echekki2015principal}
T.~Echekki, H.~Mirgolbabaei, Principal component transport in turbulent
  combustion: A posteriori analysis, Combust. Flame 162~(5) (2015) 1919--1933.

\bibitem{malik-obando-coussement-parante-2021}
M.~R. Malik, P.~O. Vega, A.~Coussement, A.~Parente, Combustion modeling using
  principal component analysis: A posteriori validation on sandia flames d, e
  and f, Proc. Combust. Inst. 38 (2021) 2635--2643.

\bibitem{10.1007/978-3-030-80542-5_23}
A.~Bellemans, M.~R. Malik, F.~Bisetti, A.~Parente, A machine-learning framework
  for plasma-assisted combustion using principal component analysis and
  gaussian process regression, in: M.~Vasile, D.~Quagliarella (Eds.), Advances
  in Uncertainty Quantification and Optimization Under Uncertainty with
  Aerospace Applications, Springer International Publishing, Cham, 2021, pp.
  379--392.

\bibitem{coussement-giquel-parante-2012}
A.~Coussement, O.~Gicquel, A.~Parente, Kernel density weighted principal
  component analysis of combustion processes, Combust. Flame 159 (2012)
  2844--2855.

\bibitem{hornik1989multilayer}
K.~Hornik, M.~Stinchcombe, H.~White, Multilayer feedforward networks are
  universal approximators, Neural networks 2~(5) (1989) 359--366.

\bibitem{de2001independent}
L.~De~Lathauwer, B.~De~Moor, J.~Vandewalle, Independent component analysis and
  (simultaneous) third-order tensor diagonalization, IEEE Transactions on
  Signal Processing 49~(10) (2001) 2262--2271.

\bibitem{anandkumar2014}
A.~Anandkumar, R.~Ge, D.~Hsu, S.~Kakade, M.~Telgarsky, Tensor decompositions
  for learning latent variable models, J. Mach. Learn. Res. 15 (2014)
  2773--2832.

\bibitem{goodfellow2016deep}
I.~Goodfellow, Y.~Bengio, A.~Courville, Deep learning, MIT press, 2016.

\bibitem{luo2012chemical}
Z.~Luo, C.~S. Yoo, E.~S. Richardson, J.~H. Chen, C.~K. Law, T.~Lu, Chemical
  explosive mode analysis for a turbulent lifted ethylene jet flame in
  highly-heated coflow, Combust. and Flame 159 (2012) 265--274.

\bibitem{Cantera}
D.~G. Goodwin, H.~K. Moffat, I.~Schoegl, R.~L. Speth, B.~W. Weber, Cantera: An
  object-oriented software toolkit for chemical kinetics, thermodynamics, and
  transport processes, \url{https://www.cantera.org}, version 2.6.0 (2022).
\newblock \href {https://doi.org/10.5281/zenodo.6387882}
  {\path{doi:10.5281/zenodo.6387882}}.

\bibitem{BansalMC2015}
G.~Bansal, A.~Mascarenhas, J.~H. Chen,
  \href{https://www.sciencedirect.com/science/article/pii/S0010218014002697}{Direct
  numerical simulations of autoignition in stratified dimethyl-ether (dme)/air
  turbulent mixtures}, Combustion and Flame 162~(3) (2015) 688--702.
\newblock \href
  {https://doi.org/https://doi.org/10.1016/j.combustflame.2014.08.021}
  {\path{doi:https://doi.org/10.1016/j.combustflame.2014.08.021}}.
\newline\urlprefix\url{https://www.sciencedirect.com/science/article/pii/S0010218014002697}

\bibitem{BhagatwalaLSSLC2015}
A.~Bhagatwala, Z.~Luo, H.~Shen, J.~A. Sutton, T.~Lu, J.~H. Chen,
  \href{https://www.sciencedirect.com/science/article/pii/S1540748914001503}{Numerical
  and experimental investigation of turbulent dme jet flames}, Proceedings of
  the Combustion Institute 35~(2) (2015) 1157--1166.
\newblock \href {https://doi.org/https://doi.org/10.1016/j.proci.2014.05.147}
  {\path{doi:https://doi.org/10.1016/j.proci.2014.05.147}}.
\newline\urlprefix\url{https://www.sciencedirect.com/science/article/pii/S1540748914001503}

\bibitem{KrismanHTBC2017}
A.~Krisman, E.~R. Hawkes, M.~Talei, A.~Bhagatwala, J.~H. Chen,
  \href{https://www.sciencedirect.com/science/article/pii/S1540748916304321}{A
  direct numerical simulation of cool-flame affected autoignition in diesel
  engine-relevant conditions}, Proceedings of the Combustion Institute 36~(3)
  (2017) 3567--3575.
\newblock \href {https://doi.org/https://doi.org/10.1016/j.proci.2016.08.043}
  {\path{doi:https://doi.org/10.1016/j.proci.2016.08.043}}.
\newline\urlprefix\url{https://www.sciencedirect.com/science/article/pii/S1540748916304321}

\bibitem{bhagatwala-hcci-dns}
A.~Bhagatwala, J.~H. Chen, T.~Lu, Direct numerical simulations of hcci/saci
  with ethanol, Combust. Flame 161 (2014) 1826--1841.

\end{thebibliography}

\end{document}